\documentclass[aps,prx,reprint,superscriptaddress,twocolumn]{revtex4-2}

\usepackage{amsmath,amssymb,amsthm,amstext}
\usepackage{graphicx}
\usepackage{color}
\usepackage{array, enumerate}
\usepackage{bm}
\usepackage{multirow}
\usepackage[breaklinks,colorlinks,citecolor=blue]{hyperref}
\usepackage{braket}
\usepackage{txfonts}
\usepackage{textcomp,gensymb}
\usepackage{enumitem}
\usepackage{xcolor}

\def\be{\begin{equation}}
\def\ee{\end{equation}}

\begin{document}
\title{Science Opportunities of Wet Extreme Mass-Ratio Inspirals}
\author{Zhenwei Lyu}
\email{zwlyu@dlut.edu.cn}
\affiliation{Leicester International Institute, Dalian University of Technology, Panjin 124221, China}
\author{Zhen Pan}
\affiliation{Tsung-Dao Lee Institute, Shanghai Jiao-Tong University, Shanghai, 520 Shengrong Road, 201210, China}
\affiliation{School of Physics \& Astronomy, Shanghai Jiao-Tong University, Shanghai, 800 Dongchuan Road, 200240,  China}
\author{Junjie Mao}
\affiliation{Department of Astronomy, Tsinghua University, Beijing 100084, China}
\author{Ning Jiang}
\affiliation{CAS Key Laboratory for Research in Galaxies and Cosmology, Department of Astronomy, University of Science and Technology of China, Hefei 230026, People’s Republic of China}
\affiliation{School of Astronomy and Space Sciences, University of Science and Technology of China, Hefei 230026, People’s Republic of China}
\author{Huan Yang}
\email{hyangdoa@tsinghua.edu.cn}
\affiliation{Department of Astronomy, Tsinghua University, Beijing 100084, China}

\begin{abstract}

Wet extreme mass-ratio inspirals (wet EMRIs), which arise from stellar-mass black holes (sBHs) inspiral into supermassive black holes (SMBHs) within the gas-rich environments of Active Galactic Nuclei (AGN), are primary sources of gravitational waves (GWs) for space-borne detectors like LISA, TianQin, and Taiji. Unlike ``dry EMRIs", which form through gravitational scattering in nuclear star clusters, wet EMRIs are naturally accompanied by interactions with accretion disks, offering rich multi-messenger science opportunities. They are distinct in generating transient electromagnetic (EM) signals, such as quasi-periodic eruptions (QPEs), which serve as valuable probes of accretion disk physics and SMBH environments. Their GW signals provide an unprecedented precision of the order of $\mathcal{O}(10^{-4}\sim 10^{-6})$ in measuring SMBH mass and spin, enabling the calibration of traditional EM techniques and offering insights into jet formation models. Additionally, wet EMRIs serve as bright and dark sirens for cosmology, facilitating percent-level precision measurements of Hubble parameter through AGN host identification or statistical association. These systems hold immense potential for advancing our understanding of black hole dynamics, accretion physics, and cosmology.

\end{abstract}
\maketitle

\section{Introduction}

Extreme mass-ratio inspirals (EMRIs) and massive black hole (BH) binaries are the two main extragalactic sources of space-borne gravitational wave detectors, such as LISA (Laser Interferometer Space Antenna) \cite{LISA_2017,Babak_2017,LISA:2022yao,LISA:2022jok,LISA:2024hlh}, TianQin \cite{Luo_2016,Luo_2020,Tianqin_2020,Gong_2021} and Taiji \cite{Gong_2021,Taiji_2017,Taiji_2020}, which operate in the millihertz frequency band. In particular, EMRIs are known for their superior precision in measuring weak environmental forces, including tidal gravitational forces from nearby stellar-mass objects \cite{Bonga:2019ycj}, disk forces associated with Active Galactic Nuclei (AGN) \cite{Kocsis2011,Yunes2011,Speri2023,Chat2023,Duque:2024mfw}, and dynamical friction from clouds of Axion-like particles populated around massive BHs \cite{Zhang:2019eid,Zhang:2018kib,Brito:2023pyl,Duque:2023seg}.

Those EMRIs formed through the interaction with accretion disks are often referred as ``Wet EMRIs" \cite{Pan:2021ksp}. It was previously pointed out in~\cite{Levin:2006uc} that star formation and evolution within AGN may lead to this class of objects. In recent years, there has been growing interest in the environmental effects influencing EMRI formation~\cite{Khalvati:2024tzz,Kejriwal:2025upp,Dyson:2025dlj,Copparoni:2025jhq,Derdzinski:2023qbi,Zwick:2021dlg,Garg:2022nko}. More recently, quantitative modeling of the nuclear star cluster and the accretion disk have shown that disk capture and migration of a stellar mass object may also give rise to rapid formation of wet EMRIs that eventually merge with the central massive BHs \cite{Pan:2021ksp}. Based on a class of the disk and cluster models, the inferred rate is comparable to or larger than those EMRIs formed through gravitational scattering within the nuclear cluster (the ``Dry EMRIs") \cite{Pan:2021oob,Pan:2021lyw}. 

In this work, we discuss several science opportunities associated with wet EMRIs. First, we propose the existence of a class of X-ray transients broadly connected to the observed quasi-periodic eruptions (QPEs \cite{Sun2013,Miniutti2019,Giustini2020,Arcodia2021,Arcodia2022,Chakraborty2021,Evans2023,Kejriwal:2024bna,Guolo2024,Arcodia2024}). If the accretion disk is inclined with respect to the spin direction of the massive BH, a stellar-mass black hole (sBH) initially migration within the disk at larger radii may exit the warped accretion disk at a smaller distance, where the GW backreaction dominates the orbital evolution. 
The sBH may collide with the warped disk twice per orbit, with ejected gas forming expanding fireballs that emit X-rays. This suggests that there is possibly a class of QPE-like transients without the association of Tidal Disruption Events (TDEs), which we refer as  type II QPEs. 

Second, we point out two promising applications of multi-messenger observation (GWs and the AGN) of wet EMRIs. It turns out that to locate wet EMRIs in their host galaxy, the LISA sensitivity limits the source redshift to $\le 0.3$. For these wet EMRIs, the measurement precision of mass and spin is generally on the order of $\mathcal{O}(10^{-4}\sim 10^{-6})$~\cite{Babak_2017}, which may serve as an accurate calibration tool for other methods of measuring BH mass and spin, such as using the broad line H$\beta$ or the X-ray spectrum. In particular, to calibrate the X-ray measurement for BH spins, we find that only sources with redshift $\le 0.1$ may have enough X-ray luminosities detectable by flag-ship missions such as Athena and eXTP. On the other hand, as GW observations can be used to measure the direction of massive BH spin, the inclination of the sBH may be used to constrain the normal direction of the accretion disk. If there is radio observation of a possible jet associated with the AGN, this multi-messenger measurement of the BH spin, jet, and disk inclination may be used to test jet formation models.

Third, wet EMRIs  offer new opportunities to study cosmology. Since we expect that wet EMRIs are associated with AGNs, searching for candidates for AGN within the error volume determined by GW measurements greatly reduces the number of candidates for wet EMRIs. This advantage applies for both bright sirens with nearby wet EMRIs (with host galaxy identification available) and dark sirens for more distant wet EMRIs. Both methods are expected to produce percent-level-precision measurement on the Hubble parameter with one-year observation of LISA. Taking into account the possibility that only a faction of AGN is observed, e.g. due to dust attenuation, does not significantly change this estimate.

These topics are by no means an exclusive list of astrophysical and cosmological applications of wet EMRIs. For example, if the surface density of the disk is sufficiently high at $\mathcal{O}(10)$ gravitational radii \cite{Kocsis2011,Yunes2011} and/or the orbital eccentricity is significant \cite{Duque:2024mfw}, the disk effect imprinted on the gravitational waveform may be detectable. Wet EMRIs may also be used to probe intermediate-mass BHs that are captured in the accretion disk \cite{Peng:2024wqf}, as these less massive objects may form mean-motion resonance pairs or chains within the AGN disk \cite{Yang:2019iqa}. In addition, the population of unresolved wet EMRIs may give rise to a stochastic GW background that is observable by space-borne GW detectors \cite{Wang:2022obu}. The scientific opportunities discussed in this work are complementary to these previous studies, which we hope will motivate further exploration in this direction.

This work is organized as follows: in Sec.~\ref{sec:population}, we discuss the formation and population of wet EMRIs, including their formation mechanisms in accretion disks and population estimates based on updated SMBH mass functions. In Sec.~\ref{sec:em-counterparts}, we explore transient electromagnetic (EM) counterparts, focusing on the connection between wet EMRIs and QPEs. In Sec.~\ref{sec:calibrate}, we address how wet EMRIs can be used to calibrate other methods of probing BH mass and spin. In Sec.~\ref{sec:test_disk_jet}, we examine the potential of wet EMRIs to test accretion disk and jet models. Finally, in Sec.~\ref{sec:cos}, we discuss the application of wet EMRIs as cosmic sirens. Resolvable host AGNs enable their use as bright sirens (yielding luminosity distance and redshift), while unresolvable hosts result in dark sirens. Finally, in Sec.~\ref{sec:conclusion}, we possible uncertainties and caveats in this analysis, and motivate further studies in this emerging field.

All masses (e.g., $M_\bullet$, $\mu$) used throughout the manuscript, including those appearing in the Fisher matrix estimates, are consistently expressed in the source frame unless explicitly stated otherwise.

\section{Multi-Messenger Science Opportunities} \label{sec:sci}
\subsection{Population estimation}\label{sec:population}
\renewcommand{\arraystretch}{1.2}
\begin{table*}[th]
\caption{Forecasted rates for stellar-mass black hole (sBH) EMRIs in the redshift range $0<z<4.5$, with the mass of sBH $\mu=10 M_\odot$. The table presents total EMRI rates, LISA detection rates (with SNR$\geq 20$) and rates for which the host AGN galaxy is resolvable. Wet EMRI models follow the configuration of our previous work~\cite{Pan:2021ksp,Pan:2021oob}, assuming a conservative AGN fraction $f_{\rm AGN}=1\%$ throughout the Universe. For comparison, the last row shows dry EMRI rates with $N_p=10$ plunges per EMRI.}
\centering
\setlength{\tabcolsep}{5pt}
\begin{tabular}{c c c | c  c  c}
\hline\hline
Wet EMRIs & AGN disk  & $T_{\rm disk}$ [yr] &  Total rates [yr$^{-1}$] &  detection rates [yr$^{-1}$] & resolvable hosts [yr$^{-1}$]\\
\hline
& $\alpha$-disk & $10^6$ & 2700 & 90 & 5\\
&  & $10^7$ & 1100 & 22 & 1.3\\
&  & $10^8$ & 390 & 9 & 0.4 \\ \cline{2-6} 
& $\beta$-disk & $10^6$ & 3500 & 100 & 6 \\ 
&  & $10^7$ & 1200 & 30 & 1\\
&  & $10^8$ & 390 & 8 & 0.6\\
\hline
Dry EMRIs & $N_p$ &   &  Total rates [yr$^{-1}$] & detectable rates [yr$^{-1}$]  & \\
\hline
& 10 &  & 480 & 32 & -\\
\hline
\end{tabular}
\label{tbl:wetEMRIs}
\end{table*}

For sBHs on inclined and/or eccentric orbits, strong interactions with the accretion disk of a rapidly accreting SMBH play a crucial role in their evolution. These interactions, mediated by dynamical friction and disk-induced density waves, damp the orbital inclination and eccentricity of the sBH relative to the disk plane and eventually capture the sBH into the disk. Once captured, the sBH begins to migrate inward, driven by the density waves it excites through its periodic motion within the disk.

For lower-mass sBHs, type-I migration dominates, where the sBH interacts with the surrounding disk material, resulting in continuous inward drift. The timescale for type-I migration is given by~\cite{Tanaka_2002,Tanaka_2004}
\begin{equation}
    t_{\text{mig,I}} \sim \frac{M}{\mu} \frac{M}{\Sigma\, r^2} \frac{h^2}{\omega_{\rm K}}\,,
\end{equation}
where $M=M(<r)$ is the total mass enclosed within the orbital radius $r$, $\mu$ is the sBH mass, $\Sigma(r)$ is the disk surface density, $h(r)=H(r)/r$ is the disk aspect ratio (with $H$ being the characteristic vertical thickness of the disk at a given radius $r$ from the central object), and $\omega_{\rm K}$ is the Kepler angular velocity. For more massive sBHs capable of opening a gap in the accretion disk, type II migration takes over. In this case, the migration is slower and is governed by the viscous evolution of the disk itself. 

The structure and dynamics of the accretion disk also influence the migration and formation rates of EMRIs. In $\alpha$-type disks, the viscosity scales with the total pressure (which includes both the gas and radiation pressure), while in $\beta$-type disks the viscosity scales with the surface density of the disk. These differences affect the disk's structure and, consequently, the migration timescales and efficiency. Migration continues until the sBH approaches the SMBH, where GW emission dominates, driving the final inspiral, as illustrated in FIG.~\ref{fig:Rscale}. 
The presence of accretion flows accelerates both the capture of sBHs into the disk and their inward migration, significantly increasing the EMRI formation rate compared to the dry channel, where no disk is present~\cite{Pan:2021oob}.

The formation and population of wet EMRIs depend on several factors, including the fraction of SMBHs hosted in AGNs, the structure of the accretion disk, and the initial distribution of sBHs in the surrounding stellar cluster. Observational studies suggest that approximately $1\% \sim 10\%$ of the SMBHs reside in AGNs, the fraction varying by redshift. For simplicity, we conservatively assume a constant AGN fraction of $f_{\text{AGN}}=1\%$ throughout the universe. 

The SMBH mass function is critical to estimate the wet EMRI population. Two representative SMBH mass functions are commonly used:
\begin{enumerate}
    \item Population III seeded model~\cite{Barausse_2012,Shankar_2008,Shankar_2013} 
    \[
    \frac{dN_{\bullet}}{d\log M_{\bullet}} = 0.01 \left( \frac{M_{\bullet}}{3 \times 10^6 M_{\odot}} \right)^{-0.3} \, \text{Mpc}^{-3}\ .
    \]
    \item Phenomenological model~\cite{Gair_2010}
    \[
    \frac{dN_{\bullet}}{d\log M_{\bullet}} = 0.002 \left( \frac{M_{\bullet}}{3 \times 10^6 M_{\odot}} \right)^{+0.3} \, \text{Mpc}^{-3}\ .
    \]
\end{enumerate}
In both cases, $dN_{\bullet} / d\log M_{\bullet}$ represents the number density of SMBHs per logarithmic mass interval. These mass functions were used in our previous studies~\cite{Pan:2021ksp,Pan:2021oob} to estimate the abundance of SMBHs in the range $10^4\sim10^7 M_\odot$, which is particularly relevant for the formation and detection of EMRIs. Our results demonstrated that wet EMRIs could significantly enhance total and detectable EMRI rates, often dominating over dry EMRIs.

Recent studies based on optical observations of TDEs have provided an updated local SMBH mass function, which is nearly flat in logarithmic space~\cite{yao2023tidal}, given by
\begin{align}\label{eq:mass_function}
    \frac{dN_\bullet}{d\,\log M_\bullet} = 0.005 \left(\frac{M_\bullet}{3\times 10^6\, M_\odot}\right)^{\,\beta} {\,\rm Mpc}^{-3}\,,
\end{align}
with a power index $\beta$ set to zero in the simulation. This updated mass function provides an empirical measurement of SMBH distribution, based on recent TDE observations \cite{yao2023tidal}. Unlike other models—such as the Population III seed model, which is derived from simulations, and the phenomenological model, which is constructed from theoretical assumptions—the TDE-based mass function is directly informed by observational data, making it potentially more reliable.

In contrast to earlier studies that extrapolated SMBH populations from higher masses ($\sim10^7 M_\odot$)~\cite{Babak_2017}, the TDE-based estimates are grounded in observations of SMBHs with masses as low as $10^5 M_\odot$~\cite{yao2023tidal}, making them particularly relevant for modeling EMRI sources in the LISA band. Although current TDE observations are limited to the local Universe ($z<0.5$), we assume this mass function to be universal and extrapolate its shape to higher redshifts ($z\sim4.5$). This extrapolation is motivated by the assumption that the fundamental black hole demographics and EMRI formation mechanisms do not evolve significantly over this redshift range.

\subsubsection{Population and detectability}
With this updated mass function, we have reevaluated the population and detection rates of wet EMRIs, as summarized in Table~\ref{tbl:wetEMRIs}. The total wet EMRI rate across all SMBHs is calculated by integrating the differential EMRI rate over the mass $M_\bullet$ and redshift $z$. The differential rate is given by 
\begin{equation}
\frac{d^2 \mathcal{R}_{\text{wet}}}{dM_{\bullet} dz} = \frac{f_{\text{AGN}}}{1+z} \frac{dN_{\bullet}}{d M_{\bullet}} \frac{dV_c(z)}{dz} C_{\text{cusp}}(M_\bullet,z)\, \Gamma_{\text{wet}}(M_{\bullet}) \,,
\end{equation}
where the factor $1/(1+z)$ arises from the cosmological redshift, $V_c(z)$ is the comoving volume of the Universe up to redshift $z$, $C_{\text{cusp}}(M_\bullet,z)$ is the fraction of SMBHs living in stellar cusps \cite{Babak:2017tow}. The time-averaged wet EMRI rate per AGN, $\Gamma_{\text{wet}}(M_{\bullet})$, depends on the disk lifetime $T_{\text{disk}}$, the rate of sBH capture, and the migration timescale \cite{Pan:2021ksp,Pan:2021lyw,Pan:2021oob}.

Table~\ref{tbl:wetEMRIs} summarizes the updated wet EMRI population estimates based on the revised SMBH mass function. The table provides the total wet EMRI rates for sBHs ($\mu=10 M_\odot$) over the redshift range $0<z<4.5$, as well as the corresponding LISA detection rates (for a signal-to-noise ratio $\mathrm{SNR}\ge 20$) and the fraction of resolvable host AGN galaxies. A host galaxy is considered resolvable if the expected number of AGNs within the localization volume is less than 1.1 (as detailed in Sec.~\ref{sec:cos}). 
These calculations are based on the framework established in our previous work~\cite{Pan:2021ksp, Pan:2021oob}, assuming a conservative AGN fraction of $f_{\rm AGN}=1\%$ throughout the Universe. For comparison, the final row includes the dry EMRI rates, assuming $N_p=10$, where $N_p$ represents the number of plunges per EMRI.

To evaluate LISA detection rates and conduct the Fisher matrix analyses presented in this work, we adopt the Augmented Analytic Kludge (AAK) waveform model~\cite{Barack:2003fp,Chua:2015mua,Chua_2017,Stein:2019buj,Chua:2018woh,Katz_2021}, as implemented in Fast EMRI Waveforms (FEW)~\cite{Katz_2021,Chua_2021,Fujita_2020,Speri:2023jte,michael_l_katz_2023_8190418}. All analyses are carried out in a 14-dimensional parameter space that includes the mass and spin of the SMBH, the mass of the sBH, the luminosity distance, the initial semi-latus rectum, initial eccentricity, sky location, orbital inclination, and the initial orbital phases in $\phi, \theta, r$. For both dry and wet EMRIs, we assume isotropic distributions for sky location and SMBH spin direction. The three phase angles are sampled uniformly from $[0,2\pi]$, and coalescence times are randomly drawn from the interval $[0,2]$ years.
For wet EMRIs, the cosine of the inclination angle is sampled uniformly from the range $[0,0.1]$, reflecting alignment with the AGN disk, while for dry EMRIs it is drawn from a uniform distribution over $[-1,1]$, allowing for isotropic orientations. The initial eccentricity $e_0$ is also assigned based on the formation channel: for dry EMRIs, $e_0$ is drawn from a uniform distribution in $[0,0.2]$, consistent with their dynamical origin in dense stellar environments. In contrast, wet EMRIs are assumed to start on circular orbits with $e_0=0$, reflecting strong eccentricity damping due to AGN disk interactions. We adopt a uniform spin distribution in the range $[0,0.99]$ for dry EMRIs, which reduces the detection rate for the dry channel by approximately $30\%$ compared to the case with fixed spin $a = 0.98$, while for wet EMRIs we assume a fixed spin of $a = 0.98$ to reflect efficient spin alignment in AGN disks.

The LISA detector's response to these waveforms is modeled using its two orthogonal channels ($I$ and $II$), with the GW signals projected as~\cite{Rubbo_2004,Barack_2004}
\begin{align}
h_I(t) &= h_+(t)\,F^+_I(t) + h_\times(t)\,F^\times_I(t)\,, \\
h_{II}(t) &= h_+(t)\,F^+_{II}(t) + h_\times(t)\,F^\times_{II}(t)\,,
\end{align}
where $h_+(t)$ and $h_\times(t)$ are the two GW polarizations, and $F^+_I(t)$, $F^\times_I(t)$, $F^+_{II}(t)$, and $F^\times_{II}(t)$ are the antenna pattern functions of the LISA detector for each channel. The SNR used to assess the detectability of EMRIs is defined as
\begin{equation}
\rho = \sqrt{\langle h_I | h_I \rangle + \langle h_{II} | h_{II} \rangle}\,,
\end{equation}
where the inner product is given by
\begin{equation}\label{eq:inner}
\langle a, b \rangle = 4 \, \text{Re} \int_{f_{\text{min}}}^{f_{\text{max}}} \frac{\tilde{a}(f) \tilde{b}^*(f)}{P_n(f)} \, df\,,
\end{equation}
where $\tilde{a}(f)$ and $\tilde{b}(f)$ are the Fourier transforms of the signals $a(t)$ and $b(t)$, and $P_n(f)$ represents the one-sided spectral noise density of the LISA detector~\cite{LISA_2017,Robson_2019,babak2021}. The detection threshold is set at $\rho = 20$~\cite{Babak_2017}. The corresponding LISA detection rates are listed in the second-to-last column of Table~\ref{tbl:wetEMRIs}.

The inclusion of the TDE-based SMBH mass function enhances our ability to model wet EMRI populations and their detectability, providing a refined picture of their contribution to the GW event rate.

\subsubsection{Identifications of wet EMRIs and host AGNs}\label{sec:Identification}
The distinction between wet and dry EMRIs can be drawn based on their orbital characteristics and potential environmental signatures in the GW signal.

In wet environments, based on standard migration theory, strong interactions between sBHs and the accretion disks of rapidly accreting SMBHs—primarily via dynamical friction and disk-induced density waves—efficiently damp the vertical momentum and eccentricity of an sBH, aligning it with the disk plane and circularizing its orbit. Consequently, wet EMRIs typically exhibit near-zero eccentricities and low inclinations by the time they enter the LISA band. 

In contrast, dry EMRIs, which form through dynamical scattering in gas-poor nuclear star clusters, where such dissipative effects are absent. These systems typically retain high eccentricities and inclined orbits throughout their evolution \cite{Pan:2021ksp,Pan:2021oob}. The capability of space-based GW detectors such as LISA to measure orbital eccentricities with high precision—for instance, down to $e\sim 10^{-5}$ \cite{Babak:2017tow,Fan_2020}—enables the potential classification of EMRIs based on their orbital morphology.

Beyond orbital shape, wet EMRIs may also be distinguished through other observational features. One such possibility arises from the influence of the AGN disk on the sBH’s orbital evolution. If the disk is sufficiently dense, the orbiting sBH can experience torques or perturbations from the surrounding gas. These interactions may imprint detectable features on the GW waveform, such as phase shifts or resonant modulations, offering direct evidence of a disk-assisted inspiral~\cite{Kocsis_2011, Yunes_2011}. Another avenue for distinction involves EM counterparts. Wet EMRIs may generate Type II QPEs — transient X-ray flares that occur when the sBH crosses the warped AGN disk twice per orbit (see Sec.~\ref{sec:em-counterparts}). The contemporaneous detection of such QPEs alongside the GW signal would provide compelling multi-messenger confirmation of a wet EMRI event. Even in the absence of a direct EM counterpart, the spatial association of an EMRI signal with a known AGN within the GW localization volume increases the probability that the event originated in a gas-rich environment.

To evaluate the potential for identifying the AGN host of an EMRI event, we utilize Fisher matrix analysis, a well-established statistical method to estimate the precision of parameter measurements for expected signals. The Fisher information matrix, $\Gamma_{ij}$, is defined as
\begin{equation}
\Gamma_{ij} = \left\langle \frac{\partial h}{\partial \lambda_i}, \frac{\partial h}{\partial \lambda_j} \right\rangle\,,    
\end{equation}
where $h$ is the signal, $\lambda_i$ are the source parameters, and the inner product is defined by Eq.~\ref{eq:inner}. The covariance matrix, $\Sigma_{ij} = (\Gamma^{-1})_{ij}$, is obtained by inverting the Fisher matrix, with the diagonal elements, $\sigma_{\lambda_i}^2 = \Sigma_{ii}$, represent the variances of the individual parameters, thereby quantifying the uncertainties and correlations associated with the source localization, including angular resolution. 

\begin{figure}[th!]
    \centering
    \includegraphics[trim={0pt 0pt 0pt 0pt},clip,width=0.98\linewidth]{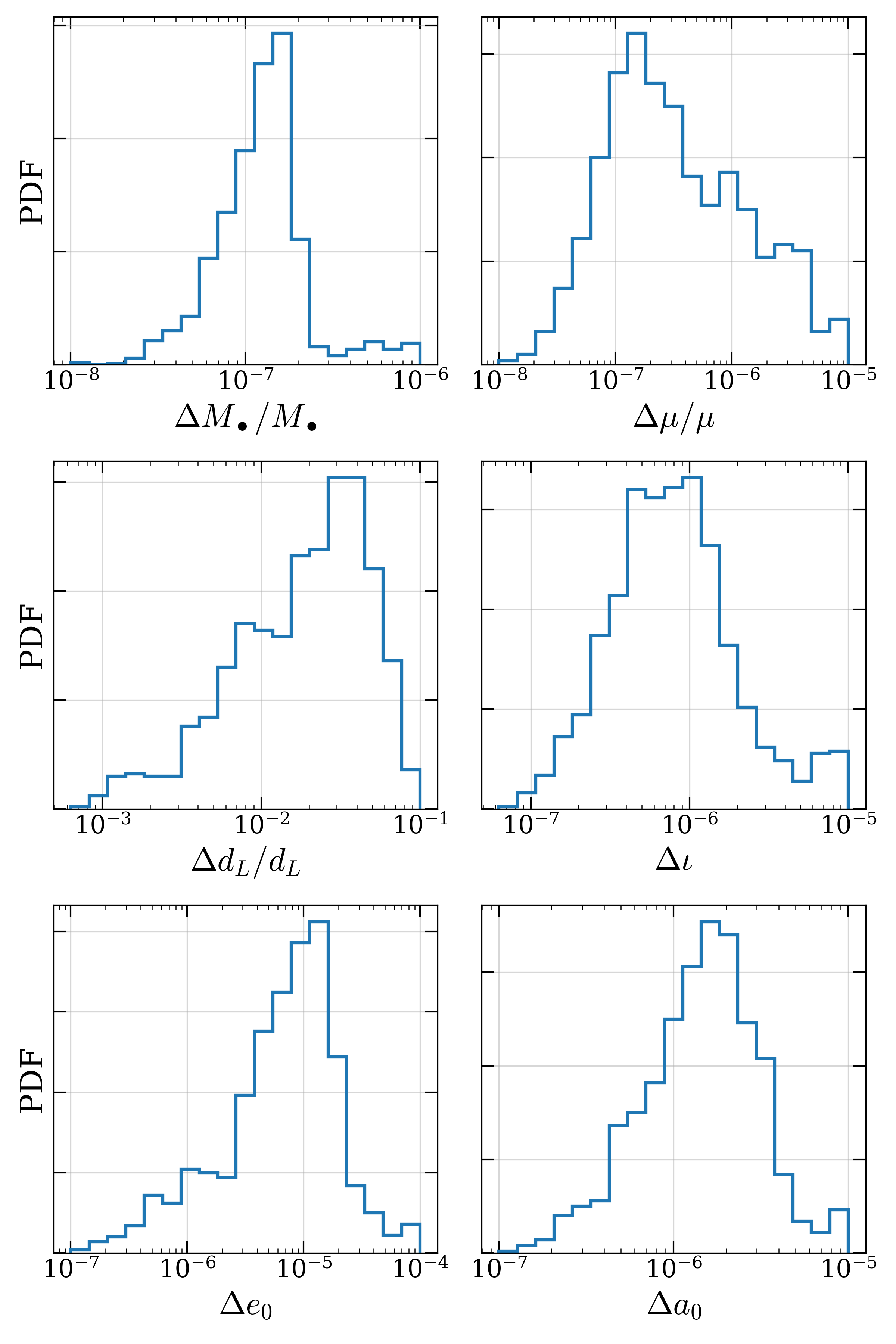}
    \caption{Histograms of $1\sigma$ parameter uncertainties from the Fisher analysis, evaluated for $\sim 2000$ EMRI events with $\mathrm{SNR} > 20$. Shown are distributions for $\Delta M_\bullet/M_\bullet$, $\Delta \mu/\mu$, $\Delta d_L/d_L$, $\Delta \iota$, $\Delta e_0$, and $\Delta a_0$. Since masses $M_\bullet$ and $\mu$ are reported in the source frame, their relative uncertainties are larger by a factor of $1+z$ compared to the detector frame.}
    \label{fig:fisher_histograms}
\end{figure}

To illustrate the expected precision in intrinsic and extrinsic parameters, we performed a Fisher matrix analysis over a population of $\sim 2000$ wet EMRI events with $\mathrm{SNR} > 20$. FIG.~\ref{fig:fisher_histograms} presents histograms of the $1\sigma$ uncertainties for the central black hole mass $M_\bullet$, compact object mass $\mu$ (source frame), luminosity distance $d_L$, orbital inclination $\iota$, initial eccentricity $e_0$, and SMBH spin $a_0$. These distributions reflect the statistical spread of parameter precisions expected in the LISA EMRI population. 

These results demonstrate that LISA will achieve high-precision measurements for both intrinsic and extrinsic EMRI parameters, consistent with prior studies (e.g., Ref.~\cite{Babak_2017}). Such precise parameter recovery will be instrumental in distinguishing EMRI formation channels and identifying host galaxies~\cite{Sun_2025} .

To compute the localization volume—the three-dimensional region in which the EMRI source is likely located—we use catalogs of wet EMRIs generated from a range of astrophysical models (see Table~\ref{tbl:wetEMRIs}). These include, for example, an $\alpha$-disk model with a characteristic disk lifetime $T_{\rm disk} = 10^6$ years. For each model, we estimate the number of potential AGN hosts located within the expected localization volumes.

The sky localization volume quantifies the three-dimensional uncertainty in the inferred source location, combining angular position and luminosity distance. In the Fisher formalism, the solid angle uncertainty $\Delta \Omega_s$ is derived from the covariance matrix $\Sigma = \Gamma^{-1}$ and is given by~\cite{Pan_2020,michael_katz_2024_10930980}
\begin{equation}\label{eq:sky2D}
    \Delta \Omega_s = 2\pi \sin{\theta_s} \sqrt{\Sigma_{\theta_s \theta_s} \Sigma_{\phi_s \phi_s} - (\Sigma_{\theta_s \phi_s})^2}\,.
\end{equation}
The full $68\%$ confidence localization volume in comoving coordinates is computed by including the correlation between angular position and luminosity distance. For a trivariate Gaussian distribution, the $68\%$ probability region corresponds to an ellipsoid with volume
\begin{equation}\label{eq:sky3D}
    V_{\text{sky}} = 28\, r^3(z)\, \sin{\theta_s}\, \sqrt{\det\left( \Sigma_{\ln d_L,\, \theta_s,\, \phi_s} \right)}\,,
\end{equation}
where $r(z)$ is the comoving distance to redshift $z$, and $\Sigma_{\ln d_L,\, \theta_s,\, \phi_s}$ is the $3 \times 3$ submatrix of the full covariance matrix corresponding to the parameters $(\ln d_L, \theta_s, \phi_s)$. The prefactor 28 arises from the volume of a 3D Gaussian ellipsoid enclosing 68\% of the total probability (see Appendix~\ref{sec:uncert} for a detailed derivation). For EMRIs detected by LISA, Fisher analysis indicates that $V_{\text{sky}}$ can be restricted to $\mathcal{O}(10^2)\, \rm Mpc^3$ at moderate redshifts ($z<0.3$~\cite{Pan_2020}), depending on the SNR and the inclusion of higher order waveform harmonics, demonstrating the high localization precision of LISA.

To assess whether an EMRI can be classified as a bright siren, we adopt an average number of galaxies per comoving volume  $\bar{n}_{\rm gal} = 10^{-2} \rm Mpc^{-3}$~\cite{Kuns_2020} and assume a conservative AGN fraction of $f_{\rm AGN} = 1\%$. Although observational studies suggest that the AGN fraction is approximately $1\%$ at low redshift and may increases to $1\%-10\%$ at higher redshifts, we adopt a constant, redshift-independent AGN fraction of $1\%$ throughout this work to ensure a conservative and robust estimate.

A system is defined as a bright siren if the expected number of AGNs within the localization volume is fewer than $1.1$. Across the various astrophysical models listed in Table~\ref{tbl:wetEMRIs}, we find that approximately $20\%\sim25\%$ of detectable wet EMRIs meet this criterion. This corresponds to an annual detection rate of 3–30 bright sirens, depending on the model and detection assumptions.

In contrast, systems with multiple AGN candidates within their localization volumes are classified as dark sirens, reflecting the ambiguity in host identification. To account for observational limitations—such as interstellar extinction  and incomplete AGN catalogs — we conservatively assume that $10\%$ of initially identified bright sirens should be reclassified as dark sirens.

\subsection{Transient Electromagnetic Counterparts: type II QPEs}\label{sec:em-counterparts}
\begin{figure}[th]
    \centering
    \includegraphics[trim={0pt 0pt 0pt 0pt},clip,width=0.98\linewidth]{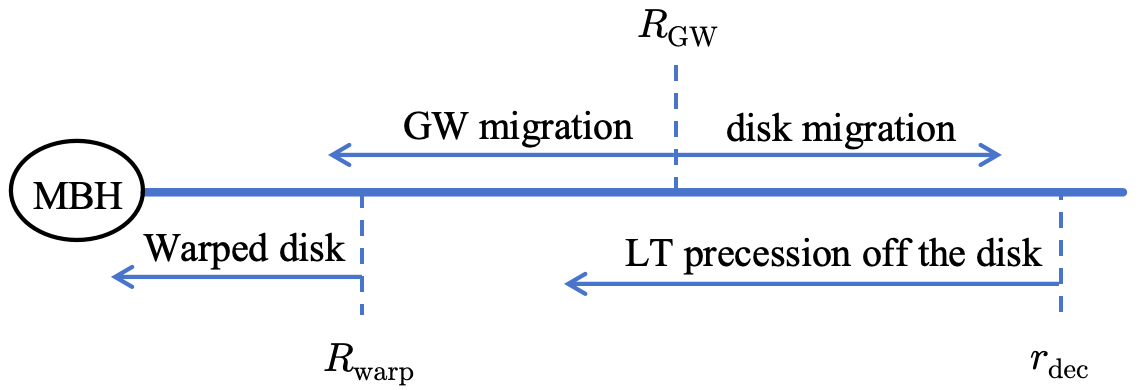}
    \caption{Different effects act at different radial scales. $R_{\rm warp}$ marks the warp radius of the accretion disk, within which the disk is warped by the relativistic Lense-Thirring effect (frame-dragging), and remains flat beyond. For $r < r_{\rm dec}$, Lense-Thirring precession dominates the orbital evolution of the sBH, potentially driving it out of the disk plane and causing misalignment. When $r < R_{\rm GW}$, GW emission governs the sBH’s migration (GW-dominated regime), while for $r > R_{\rm GW}$, migration is dominated by disk effects, such as gas torques (disk-dominated regime).
    }
    \label{fig:Rscale}
\end{figure}

In a gas-rich environment, it is natural to expect EM emissions as the sBH travels through the gas around the SMBH.
If the sBH is aligned with and embedded in the gas disk,  the continuous EM emission arising from gas accretion into the sBH will be reprocessed by the disk and will be hard to distinguish from the AGN background emission. In this case, the environmental effects on the gravitational waveform of EMRI have been investigated in previous studies \cite{Kocsis2011,Yunes2011,Speri2023,Chat2023,Duque:2024mfw}.
Notice that the disk orientation at larger radii may not be the same as the local disk direction, 
where the local disk is likely perpendicular to the SMBH spin direction \cite{Bardeen1975}, while the outer disk direction depends on the angular momentum of the gas feeding at larger radii. The warpped disk will exert  a torque onto the
central BH and align or antialign the BH spin with the angular
momentum of the outer disk. The warp radius is approximately  \cite{Natarajan1998,Lodato2013}
\be \label{eq:warp}
\frac{R_{\rm warp}}{M_\bullet} \approx 12 \left(\frac{\alpha}{0.1}\right)^{2/3} a^{2/3} \left(\frac{h}{0.1}\right)^{-4/3} \ ,
\ee 
and the disk alignment timescale in the small warp case is approximately \cite{Scheuer1996,Natarajan1998,King2005,Lodato2013,Gerosa2020}
\be 
t_{\rm align}\approx 1.2\times 10^6 \ {\rm yr} \left(\frac{\dot M_\bullet}{\dot M_{\bullet,\rm Edd}}\right)^{-1} \left(\frac{\alpha}{0.1}\right)^{5/3} a^{2/3} \left(\frac{h}{0.1}\right)^{2/3}\ ,
\ee 
where $\alpha$ is the standard viscosity coefficient in thin disks, $a$ is the dimensionless spin of the SMBH, $\dot M_{\bullet}$ is the SMBH accretion rate and the Eddington accretion rate $\dot M_{\bullet,\rm Edd} = M_\bullet/(5\times 10^7 {\rm yr})$, and $h$ is the disk aspect ratio. If the accretion of the SMBH is coherent in a period much longer than $t_{\rm align}$, we expect an equatorial accretion disk and an equatorial EMRI when the EMRI enters the LISA band; therefore, no detectable EM transient is produced.
Otherwise, if the gas feeding direction is random in each accretion episode with an accretion time scale shorter than $t_{\rm align}$,  EM transients arising from EMRI and disk collisions are possible:
for an sBH embedded in the disk, the Lense-Thirring precession tends to drive its orbit out of the disk plane,
while the disk interaction tends to keep the sBH within the plane. We denote the critical radius as $r_{\rm dec}$ (as illustrated in FIG.~\ref{fig:Rscale}),
where the alignment timescale for the sBH orbit with the disk is equal to the Lense-Thirring precession timescale of the sBH.
At smaller radii $r< r_{\rm dec}$, the sBH orbiting around the SMBH may hit the accretion disk and produce (quasi-)periodic EM transients, similar to some of the discussion of 
X-ray quasi-periodic eruptions (QPEs) \cite{Sun2013,Miniutti2019,Giustini2020,Arcodia2021,Arcodia2022,Chakraborty2021,Evans2023,Kejriwal:2024bna,Guolo2024,Arcodia2024}, where eruptions are highly likely due to collisions between a low-frequency ($f_{\rm obt}\sim 10^{-5}$ Hz) EMRI and an accretion disk formed in a recent TDE \cite{Xian2021,Linial2023,Franchini2023,Tagawa2023,Linial:2023owk,Arcodia2024,Guolo2024,Zhou2024a,Zhou2024b,Zhou2024c,Pasham2024,Chakraborty2024,linial2024coupled}. 
For clarity, we will call the TDE-associated QPEs as type I  QPEs and the 
quasi-periodic EM transients produced by an EMRI crossing an AGN disk as type II QPEs:
\begin{itemize}[noitemsep]
    \item Type I QPE: QPE associated with a  TDE
    \item Type II QPE: QPE associated with an AGN and without a TDE
\end{itemize}
Notice that Swift J023017 is likely a type II QPE source as analyzed in Ref.~\cite{Zhou2024b}, which is associated with a low luminosity AGN.

In the remainder of this section, we first estimate the decoupling radius $r_{\rm dec}$ of wet EMRIs, 
then the energy budget of type II QPEs and finally evaluate the possibility for multi-messenger detections of wet EMRIs in the LISA era.
Considering a sBH  embedded in an accretion disk which is misaligned with the SMBH equator,
the Lense-Thirring precession tends to drive the sBH out of the disk plane in a rate 
\be\label{eq:LT_prec} 
\vec \omega_{\rm LT} = \frac{2\vec{a}}{r^3}\ ,
\ee 
where  $\vec{a}$ is the dimensionless spin vector of the central SMBH and $r$ is the orbital radius.
The disk interaction tends to align the sBH's orbital plane with the disk plane at a rate $\vec \omega_{\rm DI} $ (in the direction of $\vec L \times \vec n_{\rm disk}$), where $\vec n_{\rm disk}$ is the normal vector of the disk and $\vec L$ is the angular momentum of the sBH's orbit.
The orbital angular momentum $\vec L$ evolves according to
\be \label{eq:Ldot}
\frac{d\vec L}{dt} = \left( \vec \omega_{\rm LT} + \vec \omega_{\rm DI} \right) \times \vec L\ .
\ee
Denoting the inclination of the sBH orbit relative to the disk as $\iota_{\rm sd}$,
the alignment is driven by dynamical friction at large inclinations and by density waves at small inclinations \cite{Arzamasskiy2018,Zhu2019},
with a rate 
\be 
\omega_{\rm DI} := -\frac{d\iota_{\rm sd}}{dt} =  {\rm min.}\left\{0.544 \iota_{\rm sd}, \frac{1.46h^4}{ \sin^3(\iota_{\rm sd}/2)}\right \}  \frac{1}{t_{\rm wav}}\,,
\ee 
where 
\be 
 t_{\rm wav} = \frac{M}{\mu}\frac{M}{\Sigma\, r^2}\frac{h^4}{\omega_{\rm K}}\,.
\ee 
It is straightforward to see $\omega_{\rm DI}$ reaches its maximum value
\be 
\omega_{\rm DI}^{\rm max} \approx  \frac{h}{t_{\rm wav}}\ ,
\ee  
at $\iota_{\rm sd}\approx 2h$.
The decoupling radius is determined by the equation $\omega_{\rm LT}(r_{\rm dec})=\omega_{\rm DI}^{\rm max}(r_{\rm dec})$ as 
\be 
r_{\rm dec} = 430 M_\bullet \ a^{1/8} \alpha_{0.1}^{1/8} \dot M_{\bullet,0.1}^{1/2}\mu_{30}^{-1/8} \ ,
\ee 
for $\alpha$ disks,
and 
\be 
r_{\rm dec} = 410 M_\bullet \ a^{10/59} \alpha_{0.1}^{8/59} \dot M_{\bullet,0.1}^{24/59} M_{\bullet,6}^{-2/59}\mu_{30}^{-10/59} \ ,
\ee 
for $\beta$ disks, 
where we have defined $\mu_{30} = \mu/30 M_\odot, M_{\bullet,6} = M_\bullet/10^6 M_\odot, \dot M_{\bullet,0.1} = \dot M_{\bullet}/0.1 \dot M_{\bullet, \rm Edd}$ and we have used analytic models of $\alpha$ and $\beta$ disks \cite{Kocsis2011,Yunes2011}.
The two numerical factors are nearly equal because the two disk models are similar at large radii where the gas pressure dominates the total pressure in the accretion disk.
Due to the sharp $r$ dependence of the two rates $\omega_{\rm LT}$ and $\omega_{\rm DI}$, 
the decoupling radius $r_{\rm dec}$ turns out to be insensitive to $M_\bullet, \alpha, \mu$, and only has a mild dependence on the SMBH accretion rate $\dot M_\bullet$.
From Eq.~(\ref{eq:Ldot}), it is straightforward to see that the sBH is nearly aligned with the disk ($\vec L/L \approx \vec n_{\rm disk}$) for $r > r_{\rm dec}$ 
and starts to precess around the SMBH spin axis at $r < r_{\rm dec}$ (as illustrated in FIG.~\ref{fig:Rscale}). The transition between the two regimes is very sharp due to the sharp $r$ dependence of the two rates $\omega_{\rm LT}$ and $\omega_{\rm DI}$.

As an sBH crosses an accretion disk with relative velocity $v_{\rm rel}$ higher than the local gas sound speed $v_{\rm s}$, 
the gas within the accretion radius $r_{\rm acc}:= 2G\mu/v_{\rm rel}^2$ will be shocked and the total orbital energy loss of the sBH per collision
turns out to be \cite{Zhou2024a}
\be \label{eq:delta_E}
\begin{aligned}
    \delta E_{\rm sBH} 
    &   
    = 4\pi \ln\Lambda  \frac{G^2 \mu^2}{v_{\rm rel}^2}\frac{\Sigma}{\sin(\iota_{\rm sd})}\ ,\\ 
    &\approx 2\times10^{46} {\rm ergs} \left(\frac{\ln\Lambda}{10}\right) \Sigma_5 \mu_{30}^2 r_{100} \left(\frac{\sin\iota_{\rm sd}}{0.1}\right)^{-3} \ ,
\end{aligned}
\ee 
where  $\iota_{\rm sd}$ is the angle of inclination between the EMRI orbital plane and the disk plane,
and we have defined $\Sigma_{5}=\Sigma/10^5\ {\rm g\ cm^{-3}}, r_{100}=r/100 M_\bullet$.
In the $\alpha$-disk and the $\beta$-disk models, the disk surface densities are formulated as
\be 
\Sigma(r) = 1.7\times 10^6 \ {\rm g\ cm}^{-2} \alpha_{0.1}^{-1} \dot M_{\bullet,0.1}^{-1} M_{\bullet,6}^{2} r_{100}^{3/2}\ , 
\ee 
and 
\be 
\Sigma(r) =  5.0\times 10^5 \ {\rm g\ cm}^{-2} \alpha_{0.1}^{-4/5} \dot M_{\bullet,0.1}^{3/5} M_{\bullet,6}^{1/5} r_{100}^{-3/5}\ , 
\ee 
respectively. Therefore, wet EMRIs with a warped disk are a natural mechanism to generate type II QPEs. 
If type II QPEs are confirmed in future observations, their existence provides indirect evidence for warped disks and the chaotic accretion mode in AGNs.

As the EMRI enters the LISA sensitivity band with orbital radius $r\sim 10 M_\bullet$ and orbital frequency $f_{\rm obt}\sim $ mHz,
the energy budgets per collision for the two disk models are quite different.
In the $\alpha$-disk model, it scales as $\delta E_{\rm sBH} \propto r ^{5/2} \propto f_{\rm obt}^{-5/3}$, that is to say, the mHz QPEs will be $\mathcal{O}(10^3)$ times weaker than their low-frequency counterparts.
Although the identification of mHz type II QPEs is difficult in a sea of much stronger AGN variability, the joint GW measurement of the orbits may help identify the detection of mHz type II QPEs. 
If the accretion disk can be modeled as a $\beta$-disk, the energy budget per collision scales as $\delta E_{\rm sBH} \propto r ^{2/5} \propto f_{\rm obt}^{-4/15}$, i.e. the mHz QPEs will be a few times weaker than their low-frequency counterparts. In this case, the identification of mHz QPEs is possible even without the input of GW observations. 

This multi-messenger source will be a convenient probe to the two disk models. The EMRI parameters, including the SMBH parameters $M_\bullet, a_\bullet$, the stellar-mass BH mass $\mu$ and the EMRI orbital inclination angle $\iota_{\rm sd}$, are expected to be tightly constrained from the
GW signal \cite{Babak_2017}. In combination with the AGN luminosity $L_\bullet$, it is straightforward to calculate the 
the SMBH accretion rate $\dot M_\bullet$. With these information, viscosity parameter $\alpha$ is the only unknown parameter in $\alpha$- or $\beta$-disk models, and 
the two disk models predict largely different disk surface densities at $r\lesssim 10 M_\bullet$
with a ratio
\be 
\frac{\Sigma_\alpha(r)}{\Sigma_\beta(r)} = 2.7\times 10^{-2} \alpha_{0.1}^{-1/5} \dot M_{\bullet, 0.1}^{-8/5} M_{\bullet,6}^{-1/5} r_{10}^{21/10}\ ,
\ee 
which is insensitive to the unknown parameter $\alpha$. Since the energy budget of EMRI-disk collisions is proportional to the disk surface density $\delta E_{\rm sBH}\propto \Sigma$, collisions with a $\beta$-disk naturally produce luminous QPEs that 
cannot be reconciled with the $\alpha$-disk prediction.

The other possible phenomenon to observe is the strong lensing of  flares processed by the central SMBH, similar to  self-lensing flares from SMBH binaries
\cite{Davelaar:2021eoi}. In this case, the strong lensing of the EMRI-produced flares is probably more likely to happen because the sBH is much closer to its host.
As shown in Ref.~\cite{Leong:2024nnx}, the strong lensing probability of a flare (either an EM flare or GWs) emitted at radius $r$ is
approximately
\be
P_{\rm strong\ lensing} \approx \frac{M_\bullet}{4r}\ .
\ee 
Therefore the chance of strong lensing of a single flare is around $\mathcal{O}(1\%)$ level,
and the chance will be largely enhanced considering multi-flares are produced as the EMRI orbiting around the SMBH and crossing the accretion disk at different locations. Unlike the case of SMBH binaries with one of them hosting a mini-accretion disk where the background light source is stable and the variations in the light curve mainly come from when the mini disk is lensed by the foreground BH, the light curve of mHz QPEs varies from flare to flare even without lensing. A more detailed study is required to confirm whether the lensing effect can be identified from the QPE light curve.

To summarize, wet EMRIs are promising sources of EM transients (dubbed as type II QPEs). If confirmed, they can be used for distinguishing different accretion disk models, and their existence provides indirect evidence for warped AGN disks and chaotic AGN accretion. 
In addition, wet EMRIs in the mHz band will be emitting multi-messenger signals that are possibly detectable by both spaceborne GW detectors and  X-ray telescopes,
where there is a chance that the X-ray flares get lensed by the SMBH.

\subsection{Calibrating other Methods of Probing Black Hole Mass and Spin}\label{sec:calibrate}
\begin{table}[th]
\centering
\setlength{\tabcolsep}{3pt}
\begin{tabular}{cccccccrr}
\hline\hline
\noalign{\smallskip} 
$z$ & $M_\bullet$ & ${\lambda_{\rm Edd}}$ & ${\kappa_{\rm 2-10~keV}}$ & $L_{\rm 2-10~keV}$ & $F_{\rm 2-10~keV}$  \\
 & $M_{\odot}$ &  & ${\rm erg~s^{-1}}$ &  & ${\rm erg~s^{-1}~cm^{-2}}$ \\
\noalign{\smallskip} 
\hline
\noalign{\smallskip} 
0.1 & $1.0\times10^6$ & 0.2 & 20 & $1.3\times10^{42}$ & $5.1\times10^{-14}$ \\
0.1 & $2.0\times10^6$ & 0.1 & 30 & $8.7\times10^{41}$ & $3.4\times10^{-14}$ \\
0.1 & $5.0\times10^6$ & 0.2 & 70 & $1.9\times10^{42}$ & $7.3\times10^{-14}$ \\
0.1 & $5.0\times10^6$ & 0.5 & 125 & $2.6\times10^{42}$ & $1.0\times10^{-13}$ \\
0.2 & $1.0\times10^6$ & 0.2 & 20 & $1.3\times10^{42}$ & $1.1\times10^{-14}$ \\
0.3 & $1.0\times10^6$ & 0.2 & 20 & $1.3\times10^{42}$ & $4.5\times10^{-15}$ \\
\noalign{\smallskip} 
\hline
\end{tabular}
\caption{Exemplary estimations of the $2-10$~keV X-ray luminosity and flux for SMBH counterparts at redshift $z=0.1$.
}
\label{tbl:flux_xray}
\end{table}

\renewcommand{\arraystretch}{1.4}
\begin{table*}[ht]
\caption{
Catalog of selected SMBHs with masses below $10^7 M_\odot$, whose mass and spin have been determined through EM observations. All masses are reported with $1\sigma$ errors (if provided), while spins are reported with $90\%$ error ranges. The SNRs are calculated at a fixed source redshift ($z=0.3$~\cite{Pan_2020}) using four years of data from various detectors. According to the work by Babak et al.~\cite{Babak_2017}, the measurement uncertainties for the mass and spin of the SMBH could range from approximately $10^{-6}\sim10^{-4}$ and $10^{-6}\sim10^{-3}$, respectively. 
}
\centering
\setlength{\tabcolsep}{4pt}
\begin{tabular}{c|c|c|c|c c c|c}
\hline
\hline
\multirow{2}{2.8em}{Object} & \multirow{2}{3.3em}{Mass [$\times 10^6 M_\odot$]} & \multirow{2}{2.3em}{Spin} & \multirow{2}{0.8em}{$z$} & \multicolumn{3}{c|}{SNR ($z=0.3$)} & \multirow{2}{6.5em}{Alternate\;Names} \\
 &  &  &  & LISA~\cite{Robson_2019,babak2021} & TianQin~\cite{Tianqin_2020}  & TaiJi~\cite{Taiji_2023} & \\
\hline
UGC 01032 & $\sim 1.1$  & $0.66^{+0.30}_{-0.54}$ & 0.01678 & 87 & 56 & 280 & Mrk 359 \\
\hline
UGC 12163 & $\sim 1.1$  & $>0.9$  & 0.02468 & 92 & 61 & 290 & Ark 564\\
\hline
Swift J2127.4+5654  & $\sim 1.5$  & $0.6\pm 0.2$ & 0.01400 & 66 & 38 & 210 & \\
\hline
NGC 4253 & $1.8^{+1.6}_{-1.4}$  & $>0.92$ & 0.01293 & 64 & 37 & 200 & UGC 07344, Mrk 766 \\
\hline
NGC 4051 & $1.91\pm 0.78$ & $>0.99$ & 0.00234 & 33 & 19 & 110 & UGC 07030\\
\hline
NGC 1365 & $\sim 2$  & $>0.97$ & 0.00545 & 29 & 16 & 92 &\\
\hline
1H0707-495 & $\sim 2.3$ & $>0.94$ & 0.04056 & 19 & 11 & 63 &\\
\hline
MCG-6-30-15 & $2.9^{+1.8}_{-1.6}$ & $0.91^{+0.06}_{-0.07}$ & 0.00749 & 10 & 6.8 & 36 & \\
\hline
NGC 5506 & $\sim 5$  & $0.93\pm 0.04$ & 0.00608 & 3.0 & 2.3 & 11 & Mrk 1376\\
\hline
IRAS13224-3809 & $\sim 6.3$  & $>0.975$ & 0.06579 & 1.7 & 1.5 & 6.6 &\\
\hline
Ton S180 & $\sim 8.1$ & $>0.98$ & 0.06198 & 0.9 & 0.9 & 3.7 &\\[0.5ex]
\hline
\end{tabular}
\label{tbl:EM_measure}
\end{table*}

EMRI GWs provide precise (better than $0.1\%$~\cite{Babak_2017,Fan_2020}) measurements of the mass and spin of SMBHs, which can be used to calibrate traditional EM methods if the host AGN can be identified. As discussed in the previous subsections and summarized in Table~\ref{tbl:wetEMRIs}, 
the host AGN identification are possible for a few to a few tens of EMRI host AGNs.

Generally speaking, traditional EM wave methods perform better for type I AGN than type II AGN because the latter is obscured \cite{Ricci2022,denBrok2022,MejiaRestrepo2022}. While wet EMRIs can occur in type I galaxies with $4.0<\log_{10}(M_\bullet)<7.0$, only a small fraction of galaxies in this mass range are active. Limited by available observational studies, we further limit the mass range to $5.5<\log_{10}(M_\bullet)<7.0$.

The active type I fraction of galaxies with $5.5<\log_{10}(M_\bullet)<7.0$ was estimated in a two-step process. First, following Ref.~\cite{Cho2024}, we adopt the active ($\lambda_{\rm Edd}>10^{-3}$) fraction of $\sim15\%-40\%$ for galaxies with $5.5<\log_{10}(M_\bullet)<6.5$, where $\lambda_{\rm Edd}$ is the Eddington ratio. This is broadly consistent with the active fraction of $\sim24\%-34\%$ for the same BH mass range estimated by Ref.~\cite{Gallo2010}, although a lower Eddington ratio is adopted. For a larger BH mass range $6.5<\log_{10}(M_\bullet)<10.5$, the active ($\lambda_{\rm Edd}>10^{-3}$) fraction  is $\sim10\%-16\%$~\cite{Ananna2022}. Furthermore, Ref.~\cite{Pacucci2021} yielded an active fraction of $\sim10\%-22\%$ for the host galaxy stellar mass range $9.0<\log_{10}(M_{\rm \star,~host})<10.0$. Assuming a $M_\bullet-M_{\rm \star,~host}$ relation (e.g., Eq.~39 of Ref.~\cite{Cho2024}), this translates to a BH mass range of $5.0<\log_{10}(M_\bullet)<6.0$. 

Second, the active fraction of Ref.~\cite{Cho2024} should be multiplied by the type I fraction of active galaxies. The type I fraction of active galaxies ranges from $\sim7\%$~\cite{Moran2014} to $\sim20\%$~\cite{Lu2010,Oh2015}. Therefore, the active type I fraction of galaxies with $5.5<\log_{10}(M_\bullet)<6.5$ is $\sim1\%-8\%$~\cite{Cho2024}. This is comparable to that of $\sim4\%$ for a larger BH mass range $6.5<\log_{10}(M_\bullet)<10.5$ \cite{Ananna2022}. Hence, a few percent of galaxies with $5.5<\log_{10}(M_\bullet)<7.0$ might be used to calibrate traditional X-ray measurements of the SMBH spin and optical measurements of the SMBH mass using precise GW spin and mass measurement.

For optical counterparts of such wet EMRI, the most commonly used method to estimate $M_\bullet$ is the so-called virial mass estimator, which is based on costly and time-consuming reverberation mapping or convenient yet less accurate single-epoch spectra~\cite{Shen2013,Peterson2014}. It assumes that the regions emitting broad emissions, known as broad-line region (BLR) clouds, are entirely governed by the gravitational force of the BH, so that $M_\bullet$ can be measured as long as the velocity ($v$) and distance ($r$) of the BLR clouds from the BH are known, i.e. $M_{\bullet}=f{{rv^2}\over{G}}$. There is a uniformly assigned coefficient $f$ in the calculation, which is calibrated by the $M_\bullet-\sigma_{\star}$ relation, assuming that active and inactive galaxies follow the same relation. Note that the intrinsic scatter of the tightest $M_\bullet-\sigma_{\star}$ relation for local massive galaxies is at least $0.3-0.5$ dex ~\cite{Gultekin2009,KH2013,McConnell2013}, the uncertainty associated with the $f$ used for AGNs could be considerably larger. Moreover, for a given AGN, it suffers from additional uncertainties from the kinematics, geometry, inclination of the clouds, and even the nature of the bulges~\citep{Ho2014}, making the real $f$ hugely different from object to object. Recently, some exciting progress has been made with the advanced NIR interferometry GRAVITY, mounted on the Very Large Telescope Interferometer (VLTI), which has shown promise in probing the BLR structure and thus improving the precision of the $M_\bullet$ measurement~\citep{GRAVITY2018,GRAVITY2024}. Unfortunately, it can only be applied to very few near-infrared luminous AGNs.

For X-ray counterparts of such wet EMRIs in type I AGNs, the most commonly used method to estimate BH spin is X-ray reflection spectroscopy \cite{Tanaka1995,Garcia2014}. The reflection spectroscopy takes advantage of the skewed line profile of the 6.4 keV Fe line, which is featured with a sharp blueshifted peak due to the relativistic beaming effect and an extended redshifted wing due to the gravitational redshift of matter close to the innermost stable circular orbit. More technical details, as well as descriptions of other techniques, can be found in the recent review on the observational constraints on BH spin \cite{Reynolds_2021}. 

For those active type I galaxies with wet EMRI signals, their $2-10$ keV luminosity is estimated as 
\begin{equation}
    L_{\rm 2-10~keV} = 1.3\times10^{44} {\rm ~erg~s^{-1}} \left(\frac{M_\bullet}{10^{6}~M_{\odot}}\right) \left(\frac{\lambda_{\rm Edd}}{\kappa_{\rm 2-10~keV}}\right)\ ,
\end{equation}
where $M_\bullet$ is the BH mass, $\lambda_{\rm Edd}$ the Eddington ratio, $\kappa_{2-10}~{\rm keV}$ the $2-10$ keV bolometric correction. The ratio of $\lambda_{\rm Edd}/\kappa_{\rm 2-10~keV}$ is in the range of $0.001-0.1$ \cite{Waddell2020}. Generally speaking, $\kappa_{\rm 2-10~keV}\sim20-70$ for $0.1<\lambda_{\rm Edd}<0.2$ and $\kappa_{\rm 2-10~keV}\sim70-150$ for $\lambda_{\rm Edd}>0.2$ \cite{Vasudevan2009}. The $2-10$~keV flux is obtained following the inverse square law of the luminosity distance, where we assume the standard flat $\Lambda$CDM cosmology with parameters $H_0=70~{\rm km~s^{-1}~Mpc^{-1}}$ and $\Omega_{m}=1-\Omega_{\Lambda}=0.3$. Table~\ref{tbl:flux_xray} lists exemplary estimations, mainly for $z=0.1$. The estimated $2-10$ keV X-ray flux is low in general ($\lesssim10^{-13}~{\rm erg~s^{-1}~cm^{-2}}$) but detectable. For comparisons, the XMM-COSMOS survey can reach the $2-10$~keV flux limit of $\sim9.3\times10^{-15}~{\rm erg~s^{-1}~cm^{-2}}$, with an average vignetting-corrected exposure of 40 ks \cite{Cappelluti2009}. To estimate the BH spin, merely detection is not enough. Ideally, it is preferable to match the spectral quality of the nearby AGN with the BH spin measurement \cite{Reynolds_2021}. For comparisons, the $2-10$ keV flux of Mrk~359 ($z=0.01678$, \cite{Koss2022}) listed in Table~\ref{tbl:EM_measure} is $\gtrsim 5\times10^{-12}~{\rm erg~s^{-1}~cm^{-2}}$. For the X-ray counterparts of wet EMRIs, it is essential to take deep-exposure observations using future X-ray missions (e.g., eXTP \cite{Zhang2016} and Athena \cite{Nandra2013}) with significantly larger effective areas than those of current missions.

Table~\ref{tbl:EM_measure} presents a catalog of selected SMBHs with masses below $10^7 M_\odot$, whose masses and spins have been determined using EM observations~\cite{Reynolds_2013,Reynolds_2021,ned_database}. The table also includes SNRs for GW detections from LISA, TianQin, and Taiji, assuming a source with the same intrinsic parameters placed at the redshift of $z=0.3$ and with four years of observation. Although EM observations of SMBH spin typically carry significant uncertainties due to the limitations of current methods, GW observations from wet EMRIs provide far greater precision, with uncertainties as low as $10^{-8} \sim 10^{-6}$ for mass and $10^{-7} \sim 10^{-5}$ for spin~\cite{Babak_2017,Fan_2020}, as also shown in FIG.~\ref{fig:fisher_histograms}. This comparison highlights the potential of wet EMRIs to calibrate traditional EM techniques, which rely on indirect methods such as X-ray reflection or virial mass estimators, often affected by large scatter and systematic biases. By integrating EM and GW data, Table~\ref{tbl:EM_measure} illustrates the critical role of wet EMRIs in improving the accuracy and reliability of SMBH mass and spin measurements, especially in type I AGNs.

\subsection{Testing Accretion Disk and Jet Models}\label{sec:test_disk_jet}
Astrophysical jets, such as those observed in AGN and X-ray binaries, are often believed to be launched through two primary theoretical mechanisms or their variants: the Blandford-Znajek (BZ) mechanism and the Blandford-Payne (BP) mechanism, which differ in their energy sources and jet-launching regions.

With the BZ mechanism~\cite{Blandford_1977}, a jet extracts energy and angular momentum from a spinning SMBH through magnetic fields that thread its event horizon. This process converts the rotational energy of the BH into jet power via the Penrose process~\cite{Tchekhovskoy_2011} to jet energy, so that the direction of the jet should be highly correlated with the spin axis of the BH $\hat{a}$.

In contrast, through the BP mechanism~\cite{Blandford_1982}, a jet extracts energy from the rotational motion of the accretion disk via magnetic fields that thread the disk. These fields centrifugally accelerate material along magnetic field lines and collimate it into a jet~\cite{Meier_1996}. The resulting jet aligns with the direction of the disk magnetic field, which should be highly correlated with the angular momentum $\hat{L}_{\rm disk}$. Notice that $\hat{L}_{\rm disk}$ may differ from $\hat{a}$ due to disk warping or misalignment, as discussed in Sec.~\ref{sec:em-counterparts}.

\begin{figure}[th]
    \centering
    \includegraphics[trim={0pt 0pt 0pt 0pt},clip,width=0.75\linewidth]{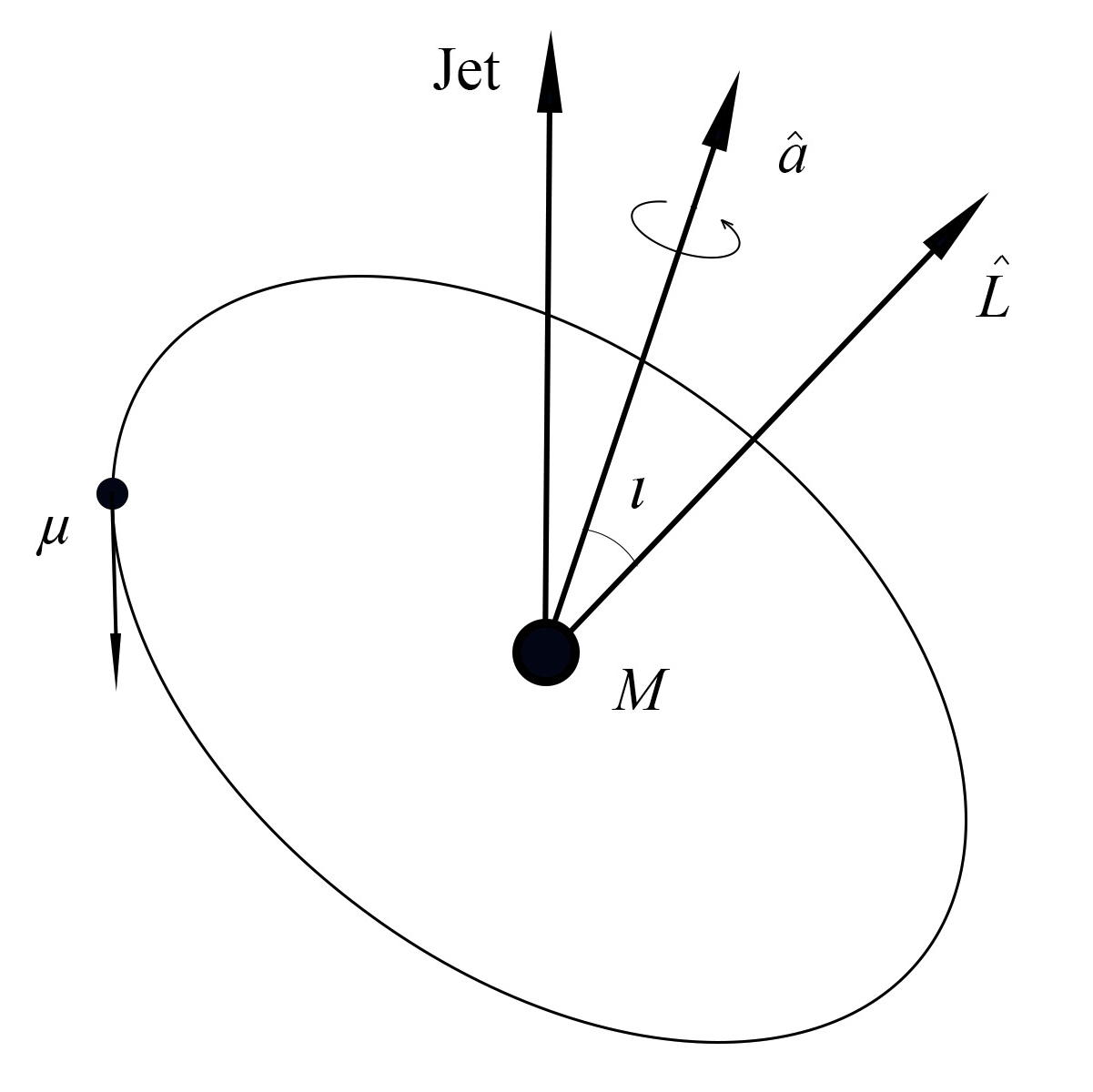}
    \caption{Relationship among the orbital angular momentum of the sBH $\hat{L}$, the SMBH spin direction $\hat{a}$, and the jet angle. The disk orientation $\hat{L}_{\rm disk}$ is inferred by tracing $\hat{L}$ backward from the LISA band ($p_0 \sim 8 M_\bullet$) to $p \sim 410 M_\bullet$ (decoupling radius), where the sBH resides within the disk.
    }
    \label{fig:three_angles}
\end{figure}

Understanding the geometric relationships among the SMBH spin axis $\hat{a}$, the accretion disk $\hat{L}_{\rm disk}$ and the jet direction (as illustrated in FIG.~\ref{fig:three_angles}) is crucial for interpreting the physical processes behind the jet-launching. Disentangling these directional vectors allows us to constrain the dominant jet-launching mechanism and to better understand the interplay between SMBH spin, disk dynamics, and magnetic fields.

\begin{figure}[th]
    \centering
    \includegraphics[trim={10pt 0pt 0pt 0pt},clip,width=0.98\linewidth]{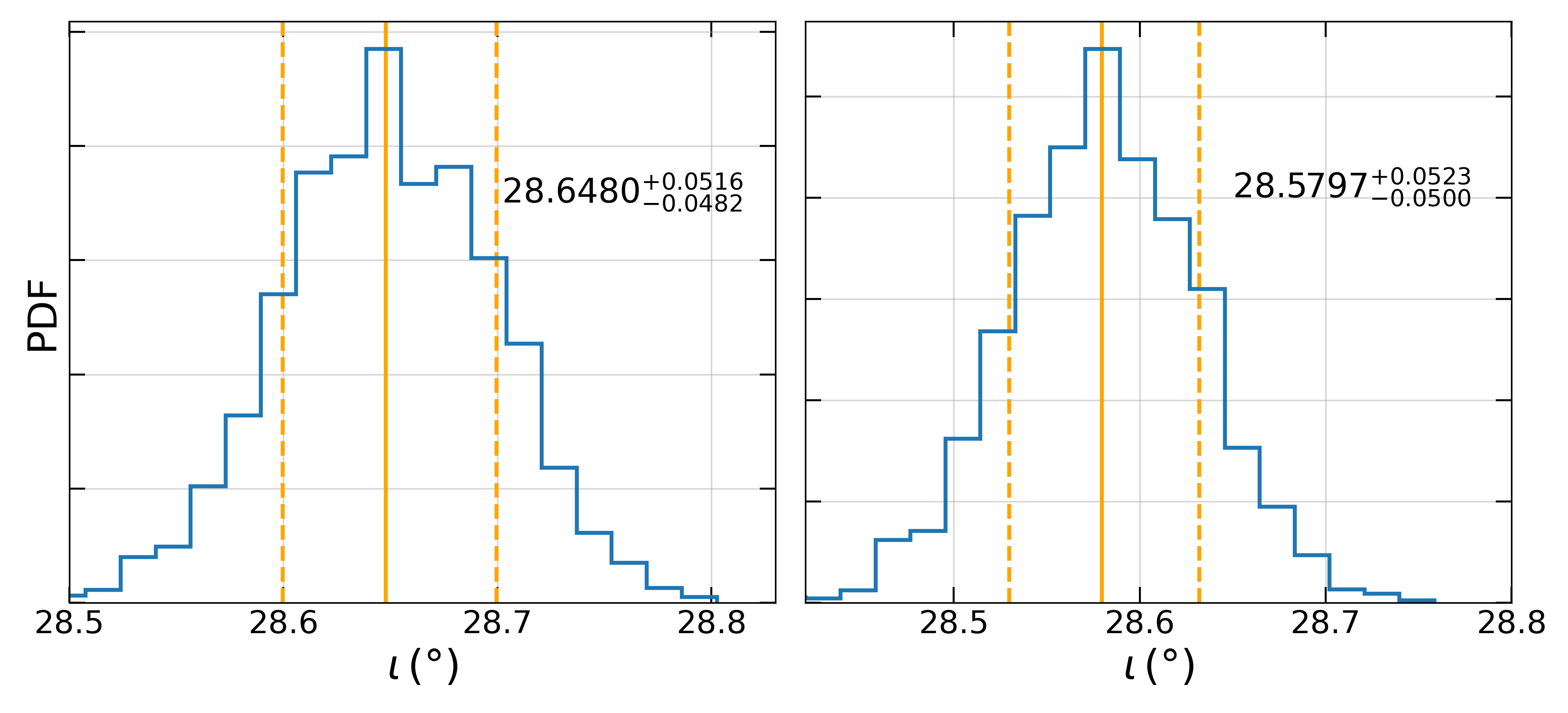}
    \caption{Posterior distribution of the inclination angle $\iota$, shown for the dimensionless semi-latus rectum at $p_0 \sim 8 M_\bullet$ (left panel), and after tracing the evolution back to $p \sim 150 M_\bullet$ due to GW radiation, and further to $p = 410 M_\bullet$ using Eq.~\ref{eq:Ldot} with $\beta$-disk model (right panel). The orange vertical lines indicate the mean value along with the $1-\sigma$ uncertainty.
    }
    \label{fig:incl_angle}
\end{figure}

\begin{figure}[th]
    \centering
    \includegraphics[trim={0pt 0pt 0pt 0pt},clip,width=0.75\linewidth]{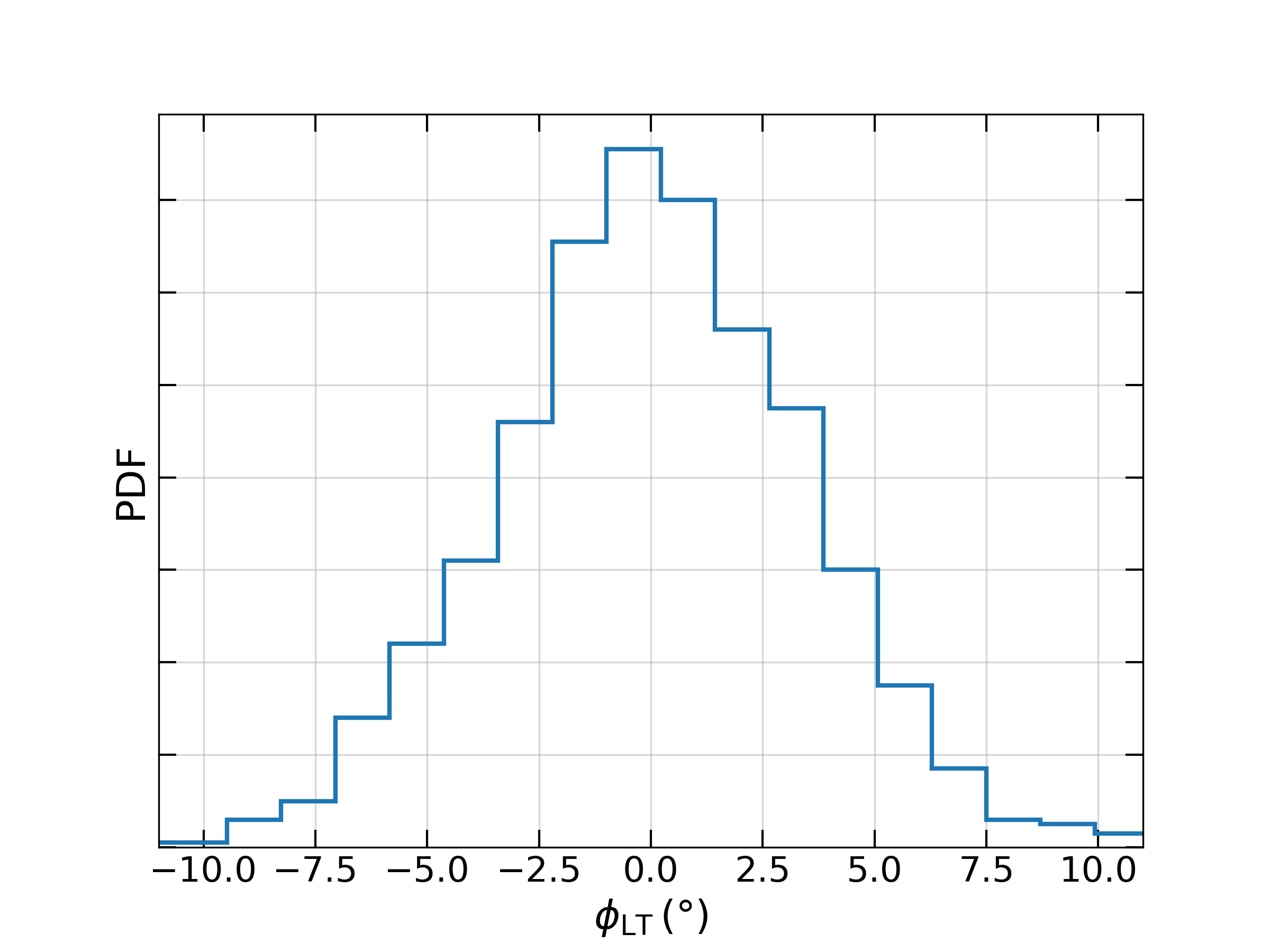}
    \caption{Uncertainty in the Lense-Thirring precession angle at $p \sim 150 M_\bullet$, traced back from a semi-latus rectum of $p_0 \sim 8 M_\bullet$.
    }
    \label{fig:LT_angle}
\end{figure}

Among the three vectors, the SMBH spin axis $\hat{a}$ is the most straightforward to measure using GW observations. The LISA mission is expected to constrain the direction of the SMBH spin with an angular uncertainty in the range of approximately $1-10 \,\rm deg^2$ for $z<0.25$~\cite{Pan_2020}. Additionally, GW signals allow accurate extraction of the inclination angle $\iota$, defined as the angle between the SMBH spin axis ($\hat{a}$) and the orbital angular momentum vector ($\hat{L}$). This measurement is typically made at a semi-latus rectum of $p \sim 8 M_\bullet$, corresponding to the region where LISA is most sensitive.

In warped disk scenarios, inferring the orientation of the accretion disk ($\hat{L}_{\rm disk}$) is more challenging, as the evolution of sBH depends on a complex interplay of forces acting at different radii (as illustrated in FIG.~\ref{fig:Rscale}). In the outer regions of the disk, the motion of the sBH is primarily governed by disk forces, while closer to the SMBH, GW radiation dominates its evolution. At larger radii, the sBH orbits within the disk plane, but as it migrates inward, it gradually departs from the disk due to Lense-Thirring precession induced by the SMBH~\cite{Bardeen:1975zz}.

To estimate the disk orientation at larger radii, we perform a backward evolution of the orbital trajectory—driven solely by GW damping and neglecting other dynamical interactions for simplicity—from the LISA band ($p \sim 8 M_\bullet$) to $p \sim 150 M_\bullet$. We then evolve the system further to $p = 410 M_\bullet$ using Eq.~\ref{eq:Ldot} with the $\beta$-disk model. Under this assumption, the orbital angular momentum vector of the sBH, reconstructed at $p \sim 410 M_\bullet$, serves as a proxy for the accretion disk orientation at that radius. This radius is significantly larger than the warp radius $R_{\rm warp}$ (see Eq.~\ref{eq:warp}) and corresponds to the outer region of the accretion disk, where disk forces are expected to dominate over relativistic effects, as discussed in Sec.~\ref{sec:em-counterparts}.

At this outer location, the disk orientation can be described by two angles measured relative to the SMBH spin axis: the inclination angle, which captures the tilt of the disk, and the precession angle, which describes its azimuthal orientation. Together, these angles define the full 3D geometry of the disk and provide essential input for understanding the jet-launching mechanism in the context of disk–black hole interactions.

The inclination angle $\iota$ of the sBH's orbit is defined as~\cite{Ryan_1995}
\begin{equation}
    \cos\iota=\frac{L_z}{\sqrt{Q+L_z^2}}\,,
\end{equation}
where $L_z$ represents the $z$-component of the orbital angular momentum of the sBH, and $Q$ is the Carter constant. The inclination angle $\iota$ represents the orientation of the orbital angular momentum relative to the spin of SMBH. However, as $\iota$ increases over time due to the radiation reaction~\cite{Hughes_2000,Glampedakis_2002,Gair_2006}, it no longer reliably indicates the orientation of the accretion disk. Thus, to estimate the disk's orientation, the orbital trajectory must be evolved backward to regions far from the SMBH, where $\iota$ can, in principle, reflect the orientation of the accretion disk.

The measurement uncertainty of $\iota$ as the EMRI system enters the LISA band ($p_0 \sim 8 M_\bullet$) is estimated by performing Fisher analysis with the AAK waveform model~\cite{Chua_2017,Katz_2021}. The posterior samples obtained from this analysis are propagated backward by evolving the orbital trajectory first to $p \sim 150 M_\bullet$, and then to $p\sim 410M_\bullet$.  At this larger radius, the orbital angular momentum of the sBH serves as a proxy for the orientation of the accretion disk. As an illustrative example, we consider a simulated EMRI system with the following key parameters:
\begin{itemize}[noitemsep]
    \item SMBH mass: $M_\bullet=10^6M_\odot$
    \item SMBH spin: $a=0.98$
    \item sBH mass: $\mu=10M_\odot$
    \item initial eccentricity: $e=0.01$
    \item observation time: $T=2$ years
    \item inclination angle: $\iota\approx 30^\circ$
    \item SNR: fixed at $50$ by adjusting the luminosity distance
\end{itemize}

The initial eccentricity of the system is set to $e = 0.01$, consistent with expectations for disk-driven (wet) EMRIs. While idealized wet EMRI models often assume perfectly circular orbits ($e=0$), recent studies suggest that AGN disk interactions do not fully circularize the orbit. In particular, turbulence and multi-body resonances within the disk can induce residual eccentricities in the range $e \sim 10^{-4}$ to $10^{-2}$ in the LISA band, depending on specific disk properties such as the surface density and coherence time~\cite{Sun_2025}. The choice of $e = 0.01$ thus reflects a realistic and observationally motivated value for wet EMRIs, rather than a purely idealized scenario.

Furthermore, this residual eccentricity has implications for parameter estimation and waveform modeling. As shown in FIG.~\ref{fig:fisher_histograms}, LISA's eccentricity measurement precision is expected to reach $\sim 10^{-5}$, enabling the detection of such small but nonzero values. In this work, we do not incorporate eccentricity when propagating posterior samples of the inclination angle $\iota$ to larger radii (e.g., $p \sim 150$ and $410\,M_\bullet$) for estimating uncertainties in disk geometry. The eccentricity contributes weakly to this propagation compared to inclination and spin, but its inclusion ensures a more realistic assessment of EMRI dynamics in AGN environments.

We find that the $1\sigma$ uncertainty in $\iota$ at $p\sim 410M_\bullet$ is approximately $\sim 0.05^\circ$ (see FIG.~\ref{fig:incl_angle}). This result demonstrates that GW observations can provide a highly precise estimate of the disk inclination.

To estimate the uncertainty in the precession angle, we similarly propagate samples of orbital configurations measured at $p_0 \sim 8 M_\bullet$ (within the LISA band) backward to a semi-latus rectum of $p \sim 150 M_\bullet$. Ideally, the backward propagation should extend to $p\sim 400M_\bullet$ as illustrated in FIG.~\ref{fig:Rscale}. However, for $p>150M_\bullet$, the disk begins to dominate the orbital evolution. Beyond this point, the uncertainty no longer improves significantly due to the increasing influence of disk-induced torques. At these outer radii, the precession angle reflects the azimuthal orientation of the disk. However, unlike the inclination, the uncertainty in the precession angle grows approximately linearly with distance, reaching a value of around $10^\circ$ already at $p \sim 150 M_\bullet$, as shown in FIG.~\ref{fig:LT_angle}. This level of uncertainty makes it challenging to resolve the azimuthal orientation of the warped disk.

In addition to GW-derived measurements of the SMBH spin and the inclination angle, EM observations can estimate the jet direction if the host AGN of the EMRI is resolvable. For AGNs with small jet viewing angles ($\sim 10^\circ$), radio observations, such as those conducted by the MOJAVE program, can determine the jet angle with an accuracy of approximately $1^\circ$ by analyzing jet kinematics~\cite{Hovatta_2008,Pushkarev_2009,Lister_2009,Lister_2016,Pushkarev_2017,Lister_2018,Lister_2019,Lister_2021}. Furthermore, for gamma-ray bright AGNs, observations from {\it Fermi}-LAT typically measure larger jet viewing angles and greater accuracy on average compared to non-LAT-detected AGNs~\cite{Pushkarev_2009,Pushkarev_2017}.

The ability to measure the directions of the SMBH spin, accretion disk, and jet opens a promising avenue for testing jet-launching models. As illustrated in FIG.~\ref{fig:three_angles}, GW observations provide precise measurements of the spin axis and the disk inclination angle, while EM observations reveal the jet orientation. However, due to the uncertainty in the precession angle, the relative angle between the jet direction and the accretion disk remains inaccessible in individual events. As a result, while we can constrain the angle between the jet and the SMBH spin, which is most relevant for testing the BZ mechanism, we cannot yet directly compare the jet to the disk orientation, limiting our ability to test the BP mechanism on a case-by-case basis. In addition, to statistically establish the correlations, it will likely require studying a population of wet EMRIs (those with significant disk interactions) with well-resolved EM counterparts.

\subsection{Cosmology with Wet EMRIs}\label{sec:cos}
Wet EMRIs provide a promising avenue for probing cosmological models by serving as both ``bright" and ``dark" sirens. These systems combine GW measurements of the luminosity distance with EM observations to determine the redshift of their possible host galaxies. When the host galaxy can be unambiguously identified, wet EMRIs act as bright sirens, enabling an independent measurement of the Hubble parameter. However, to qualify as a bright siren, a wet EMRI requires high localization precision and a unique association with its host AGN. 
In practice, selection effects including galaxy survey incompleteness and AGN luminosity function need to be taken into account  
in a self-consistent way, e.g.,  a Bayesian approach that has been widely used for population inference in previous GW cosmology works \cite{Chen:2017rfc,Zhu:2024qpp}.
As a proof of concept, we idealize the host identification by assuming a complete galaxy survey where galaxies are either detectable 
AGNs or others (non-detectable AGNs or normal galaxies).
With this simplification, the host AGN identification is established as long as the EMRI error volume is sufficiently small containing only one AGN. As shown by forecast in Sec.~\ref{sec:Identification}, only low-redshift  ($z\lesssim 0.3$) host AGNs can be identified 
with EMRI GW signals. In the following forecast of host AGN identification, we adopt a conservative assumption that $1\%$ of low-redshift galaxies are detectable AGNs (see Sec.~\ref{sec:calibrate} for a brief summary of AGN detection).

Dark sirens, despite the absence of unambiguous host galaxy identification, still play a significant role in measuring the Hubble parameter. In such cases, statistical methods can be employed, utilizing galaxy catalogs to infer redshift distributions within the localization volume. Although this approach offers reduced precision compared to bright sirens, it enables dark sirens to complement bright sirens in constraining cosmological parameters, as dark sirens primarily use EMRIs with higher redshift than bright sirens. In addition, to account for observational challenges, such as interstellar extinction or incomplete AGN catalogs, we conservatively reclassify $10\%$ of the initially identified bright sirens as dark sirens.

In summary, wet EMRIs offer a unique opportunity to probe the Hubble parameter through bright and dark sirens. Bright sirens provide precise, independent measurements when host galaxies are resolvable, while dark sirens contribute through statistical inference methods. The following sections will explore the respective roles of bright and dark sirens in constraining the Hubble parameter.

\subsubsection{Wet EMRIs as bright sirens}

The groundbreaking detection of GW170817, with its well-identified host galaxy (NGC 4993) and associated kilonova AT 2017gfo, marked the first GW-based measurement of the Hubble constant, yielding $H_0 = 70.0^{+12}_{-8}\, \rm km\, s^{-1}\,Mpc^{-1}$~\cite{GW170817_Hubble}. This pioneering result demonstrated the potential of joint GW and EM observations to independently constrain cosmological parameters. Building on this framework, wet EMRIs represent another promising class of bright sirens, particularly in the context of future cosmological studies.

Wet EMRIs, which involve compact objects spiraling into SMBHs within AGN disks, are ideal candidates for joint GW-EM observations. Similar to past events such as GW170817, wet EMRIs can act as bright sirens if their host AGNs are uniquely identified, allowing for precise redshift measurements. In addition, wet EMRIs have several advantages over other sources, such as binary neutron star (BNS) or neutron star–black hole mergers. These advantages include higher expected detection rates, the ability to probe greater cosmological distances, and robust EM counterparts.
Their relatively high localization precision and strong association with AGNs allow for a confident identification of host galaxies and their corresponding redshifts, making them powerful tools for constraining $H_0$ and other cosmological parameters at cosmological scales.

To measure $H_0$ using wet EMRI bright sirens, the luminosity distance $d_{\rm L}$ obtained from GW signals is combined with the redshift $z$ of the host AGN. Redshift uncertainties from spectroscopic measurements are typically in the range $z\sim 0.001-0.01$, consistent with forecasts from DESI~\cite{DESI_2024} and LSST~\cite{LSST_2009}. Since almost all host-identified EMRIs (bright sirens) are expected to lie at $z<0.3$, where spectroscopic redshifts are more accurate, the influence of redshift uncertainty is negligible for our analysis.

Assuming a flat $\Lambda$CDM cosmology, the relationship between luminosity distance and redshift ($d_{\rm L}-z$ relation) is expressed as
\begin{equation}
    d_{\rm L}(z) = c(1+z) \int_0^z \frac{dz'}{H(z')}\ ,
\end{equation}
where the Hubble parameter, $H(z)$, is given by
\begin{equation}
    H(z) = H_0 \sqrt{\Omega_{\rm m,0}(1+z)^3 + \Omega_{\rm \Lambda,0}}\ ,
\end{equation}
and $\Omega_{\rm m,0}$ and $\Omega_{\rm \Lambda,0}$ are the present-day matter and dark energy densities, respectively, satisfying $\Omega_{\rm m,0} + \Omega_{\rm \Lambda,0} = 1$. The present value of the Hubble parameter, $H_0$, can be expressed in terms of the dimensionless parameter $h$, such that $H_0 = h \times 100 \, \rm km \, s^{-1} \, Mpc^{-1}$.

\begin{figure}[th]
    \centering
    \includegraphics[trim={0pt 0pt 0pt 0pt},clip,width=0.98\linewidth]{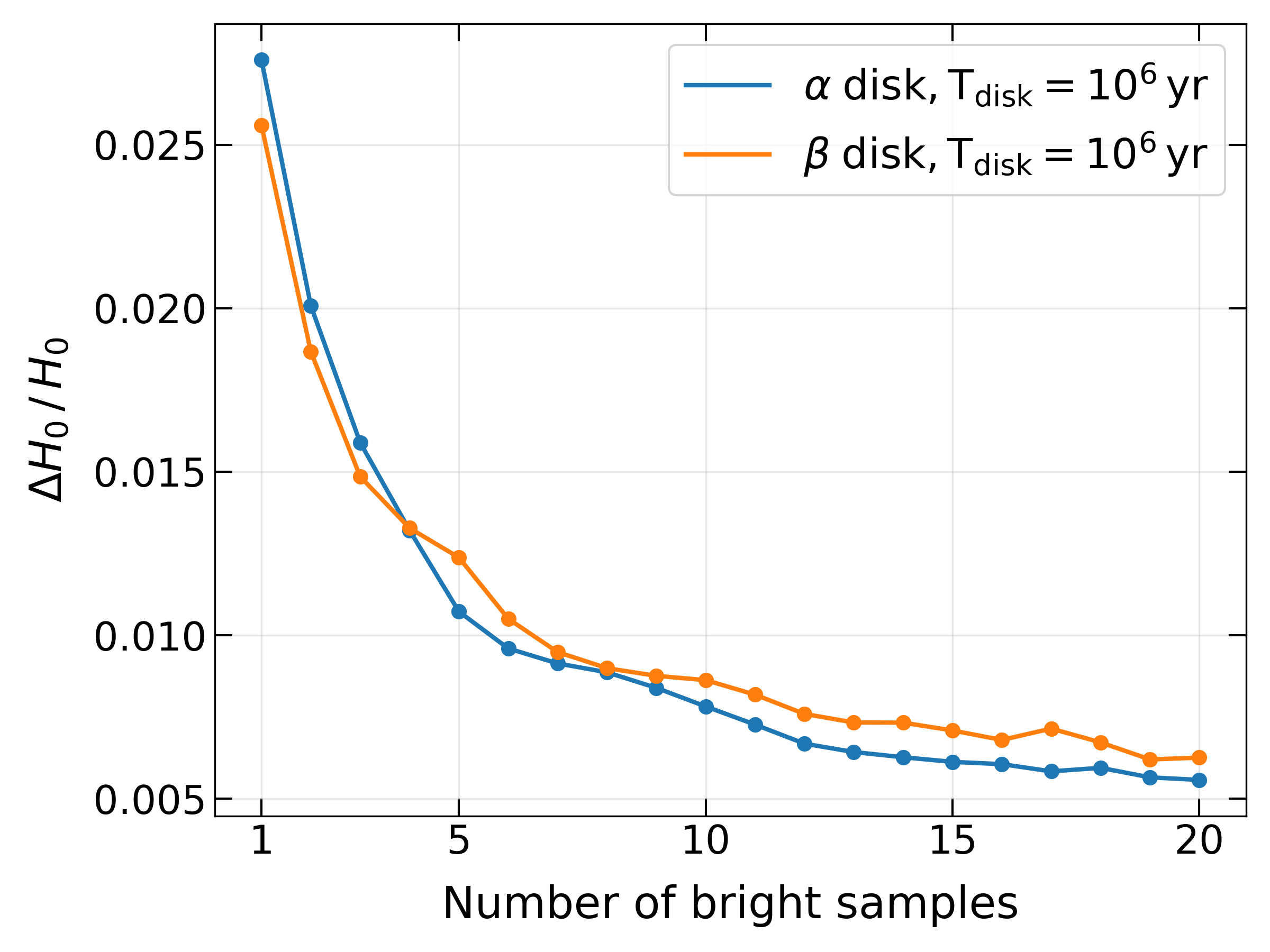}
    \caption{1-$\sigma$ relative uncertainty of the Hubble parameter $H_0$ as a function of the number of bright EMRIs is shown for two representative wet EMRI models. As expected, the uncertainty on $H_0$ decreases with increasing numbers of bright EMRIs, reaching $1\%$ with seven bright sirens. Other simulated wet EMRI models exhibit similar behavior. The colored dots represent the calculated data points}
    \label{fig:bright}
\end{figure}

For each EMRI model listed in Table~\ref{tbl:wetEMRIs}, the Hubble parameter $H_0$ can be constrained by statistically combining the luminosity distance derived from GW, $d_{\rm L}$, and its associated uncertainty, $\sigma(d_{\rm L})$, with the redshift of the host AGN. To achieve this, we randomly sample $N$ bright sirens from the catalog corresponding to the chosen EMRI model and perform Markov-Chain Monte Carlo (MCMC) simulations to derive the posterior distributions of dimensionless Hubble parameter $h$ and other cosmological parameters, such as matter density $\Omega_{\rm m,0}$.

To reduce the effects of randomness in the sampling process, this procedure is repeated 90 times for each value of $N$. By averaging the results across these repetitions, we estimate the mean value of $H_0$ and its corresponding uncertainty, $\Delta H_0$, ensuring statistical robustness. This approach ensures that the constraints derived from $H_0$ are reliable and representative of the underlying EMRI population for the selected astrophysical model.

FIG.~\ref{fig:bright} shows the 1-$\sigma$ relative uncertainty in $H_0$ as a function of the number of bright EMRIs, based on two representative wet EMRI models. As expected, the uncertainty on $H_0$ decreases as the number of bright EMRIs increases. The results suggest that with approximately seven bright EMRI detections, $H_0$ can be restricted to within $1\%$ uncertainty. This level of precision is comparable to that achieved using EM standard candles, such as Type Ia supernovae~\cite{Riess_2021}, and has the potential to help resolve the ``Hubble tension" between early- and late-universe $H_0$ measurements. Furthermore, the higher redshifts accessible to wet EMRIs provide an opportunity to probe additional cosmological parameters, such as the matter density $\Omega_{\rm m,0}$ and the dark energy equation of state parameter.

The inclusion of wet EMRI bright sirens in cosmological studies enhances the scope of GW-based standard siren measurements, complementing those from massive black hole binaries (MBHBs), BHSs, and binary black holes (BBHs). In parallel, MBHBs detected by space-based interferometers like LISA also serve as powerful bright sirens. As demonstrated by Tamanini et al.~\cite{Tamanini_2016}, LISA can constrain $H_0$ to sub-percent precision, particularly in its most optimistic configurations, with uncertainties reaching as low $0.5\%$. Importantly, MBHB standard sirens extend to redshifts as high as $z\sim8$, significantly beyond the range accessible to wet EMRIs and EM probes. This broader redshift coverage allows MBHBs to provide independent and complementary constraints on cosmological parameters.

The inclusion of wet EMRI bright sirens in cosmological studies enhances the scope of GW-based standard siren measurements, complementing those from BNS and BBH mergers. Future multimessenger detections will enable combined constraints from bright sirens across various GW source classes, significantly refining measurements of $H_0$ and other cosmological parameters. This, in turn, will deepen our understanding of the expansion history of the Universe and the underlying physics of dark energy.

\subsubsection{Wet EMRIs as dark sirens}
The dark siren analysis technique is a powerful statistical method for inferring cosmological parameters from GW events that lack identifiable EM counterparts. Instead of requiring a uniquely resolved host galaxy, this approach utilizes the spatial localization of the GW source and statistically matches it with the galaxy or AGN catalogs~\cite{Laghi:2021pqk,cosmolisa}. By assigning probabilities to potential host galaxies within the GW localization volume, the technique extracts redshift information and combines it with the GW-derived luminosity distance to constrain parameters such as the Hubble parameter $H_0$. This method is particularly well-suited for EMRIs or BBH mergers that lack direct observable EM counterparts.

The technique operates within the framework of Bayesian inference. For a set of GW events $D = \{D_1, \dots, D_N\}$, the posterior distribution of the cosmological parameters $\mathbf{\Theta} = \{H_0, \Omega_{\rm m,0}, \dots\}$ is given by
\begin{equation}
    p(\mathbf{\Theta} \,|\, D, \mathcal{H}, I) = \frac{p(D \,|\, \mathbf{\Theta}, \mathcal{H}, I)\, p(\mathbf{\Theta} \,|\, \mathcal{H}, I)}{p(D \,|\,\mathcal{H}, I)}\,,
\end{equation}
where $\mathcal{H}$ is the assumed cosmological model, $I$ represents prior information, and $p(\mathbf{\Theta} \,|\, \mathcal{H}, I)$ is the prior on the parameters. The evidence $p(D \,|\, \mathcal{H}, I)$ serves as a normalization constant, while the likelihood $p(D \,|\, \mathbf{\Theta}, \mathcal{H}, I)$ encodes the probability of the data given the parameters. For a single GW event, the likelihood is expressed as
\begin{align}
    p(D_i \,|\, \mathbf{\Theta}, \mathcal{H}, I) =& \int \int \, dz_{\rm gw} \, dd_L\, p(d_L \,|\, z_{\rm gw}, \mathbf{\Theta}, \mathcal{H}, I) \,\nonumber\\ 
    & p(z_{\rm gw} \,|\, \mathbf{\Theta}, \mathcal{H}, I) \; p(D_i \,|\, d_L, z_{\rm gw}, \mathbf{\Theta}, \mathcal{H}, I)\,,
\end{align}
where $d_L$ is the luminosity distance, $z_{\rm gw}$ is the redshift of the GW source, and $p(d_L \,|\, z_{\rm gw}, \mathbf{\Theta}, \mathcal{H}, I) = \delta(d_L - d(z_{\rm gw}, \mathbf{\Theta}))$ relates the luminosity distance to the redshift through the assumed cosmological model. The term $p(z_{\rm gw} \,|\, \mathbf{\Theta}, \mathcal{H}, I)$ represents the prior probability of the source redshift, which is informed by the galaxy catalog. The final term, $p(D_i \,|\, d_L, z_{\rm gw}, \mathbf{\Theta}, \mathcal{H}, I)$, is the likelihood of GW data, which is typically modeled as a Gaussian centered on the observed $d_L$ with uncertainties from GW parameter estimation and weak lensing~\cite{Tamanini_2016}.

Although the general dark-siren method applies to all GW events, wet EMRIs offer a distinct advantage due to their strong association with AGNs. For BBH mergers, all galaxies within the error volume of a single GW event must be considered as potential hosts. In contrast, for wet EMRIs, the search is limited to galaxies that host AGNs (we conservatively take an AGN fraction $f_{\rm AGN}=1\%$), significantly reducing the number of candidates. This reduction improves the statistical association between the GW source and its host galaxy, leading to tighter constraints on cosmological parameters. 

To statistically associate host galaxies with the GW source, the redshift prior $p(z_{\rm gw} \,|\, \mathbf{\Theta}, \mathcal{H}, I)$ is constructed by summing over all potential hosts -- specifically, the host AGN galaxies in this study -- within the GW localization volume, as follows:
\begin{equation}
    p(z_{\rm gw} \,|\, \mathbf{\Theta}, \mathcal{H}, I) \propto \sum_{j=1}^{N_{\rm g}} w_j \exp\left(-\frac{(z_j - z_{\rm gw})^2}{2\sigma_{z,j}^2}\right)\ ,
\end{equation}
where $N_{\rm g}$ is the number of candidate host galaxies, $z_j$ is the catalog redshift of the $j$-th galaxy, $\sigma_{z,j}$ includes uncertainties from peculiar velocities and $w_j$ is the weight assigned to each galaxy based on its spatial position to the GW sky localization.

As summarized in Table~\ref{tbl:wetEMRIs}, the wet EMRI models used predict the detection of 10–150 events per year, and most of the host AGNs are still poorly identified. For dark siren analysis, we adopt the first model (a $\alpha$-disk model with $T_{\rm disk} = 10^6$ years) as an example. This model predicts approximately 100 wet EMRIs per year with unresolvable host AGNs. In this simulation, we randomly select 30 dark EMRIs from the first model’s catalog to constrain the Hubble parameter $H_0$. The sample size allows us to study a representative sample of the dark EMRI population while maintaining computational efficiency. 

\begin{figure}[th]
    \centering
    \includegraphics[trim={0pt 0pt 0pt 0pt},clip,width=0.98\linewidth]{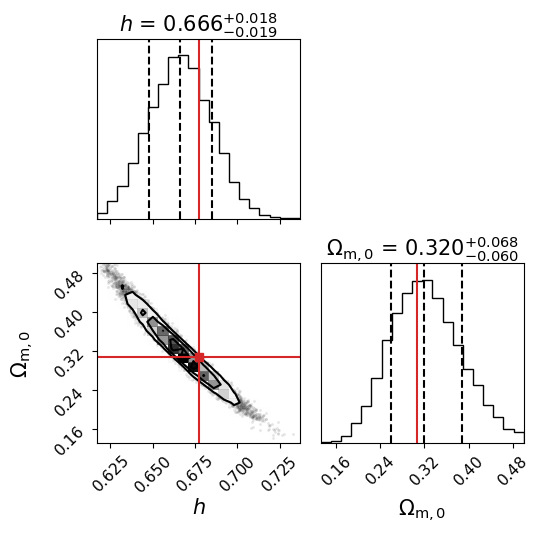}
    \caption{Posterior distributions of the dimensionless Hubble parameter $h$ and matter density $\Omega_{\rm m,0}$ inferred from 30 wet EMRI dark sirens randomly selected from the first $\alpha$-disk model. The solid red lines represent the true values, while the dashed lines indicate the mean and 1-$\sigma$ standard deviation of the distributions. }
    \label{fig:dark}
\end{figure}

Each dark EMRI event is matched with an AGN catalog, using the reduced number of AGNs within the GW localization volume to build a redshift prior to dark siren analysis. The GW localization volume of each dark EMRI typically contains $\mathcal{O}(10^2) \sim \mathcal{O}(10^3)$ potential host galaxies. However, considering that only about $1\%$ of galaxies host AGNs, the pool of candidate AGNs is reduced to $\mathcal{O}(10) \sim \mathcal{O}(10^2)$. This reduction significantly improves the statistical association between the source of GW and its host AGN compared to dry EMRIs, resulting in tighter constraints on $H_0$ and other cosmological parameters. By statistically matching the 30 selected dark EMRIs with AGN catalogs, the Hubble parameter can be constrained to $3\%$ precision, as shown in FIG.~\ref{fig:dark}. This level of precision, while slightly less than that achieved with bright sirens, underscores the power of dark sirens for independent cosmological measurements. 

This method offers a promising pathway for advancing GW cosmology, particularly in scenarios where EM counterparts are absent. Furthermore, the combination of bright and dark sirens provides complementary constraints across different redshift ranges: bright sirens probe lower redshifts with uniquely identifiable hosts, while dark sirens extend measurements to higher redshifts using statistical associations. By combining the complementary strengths of bright and dark sirens, GW cosmology offers a robust and independent means of investigating the Universe’s expansion history, leading to a more comprehensive understanding of its evolution.

\section{Discussion and Conclusion}\label{sec:conclusion}
The various science opportunities discussed in Sec.~\ref{sec:sci} are still subject to several caveats and uncertainties, two of which are highlighted below.

First, there are still theoretical uncertainties regarding the profile of the disk and the migration of stars and stellar-mass objects within the disk. Previously, most work assumed the thin-disk profile, although at larger radii additional heating from stars is included to balance out the gravitational instabilites \cite{Thompson:2005mf}. However, it is still unclear how many embedded objects are present within the disk and how they modify the structure of the disk.  Between the GW dominated region and the Broad-Line region, some work has suggested that the disk becomes clumpy and rather turbulent, which further affect both the migration of stellar-mass objects within the disk and the rate of capturing from the nuclear star cluster. Although eventually the quasi-equilibrium wet EMRI formation is mainly determined by the supply rate of sBHs to the nuclear star cluster \cite{Pan:2021lyw} rather than by detailed migration dynamics, it is necessary to understand these underlying physical processes before one can make further quantitative predictions.

Second, in Sec.~\ref{sec:sci} we have made the implicit assumption that wet EMRIs are associated with {\it visible} AGNs. We have tried to include an estimation of the ``dark" sources because of dust absorption, especially in the Hubble parameter measurement. There is, however, another scenario that the EMRI is brought to the vicinity of the massive BH because of disk migration, but the disk becomes quiescent during the time that this EMRI is observed by space-borne GW detectors. This scenario mainly affects wet EMRIs formed at the end of disk life cycles. If the lifetime of the disk is on average $T_{\rm disk}$ and the wet EMRI formation rate for that disk-assisted system is approximately once per $T_{\rm EMRI}$ (see the discussion in Ref.~\cite{Pan:2021ksp}), then this effect may affect a fraction of $T_{\rm EMRI}/T_{\rm disk}$ wet EMRIs. A more detailed study of this effect may include the time-dependent formation rate of wet EMRIs as the system starts to deplete sBHs within the nuclear star cluster.

Despite these uncertainties, the multi-messenger aspects of wet EMRIs are still less explored, partly because their abundance has only started to be quantitatively studied in recent years~\cite{Pan:2021ksp,Pan:2021oob,Pan:2021lyw}. The applications in EM transients, in calibrating disk mass and spin measurements, in testing disk and jet models, and in cosmology discussed in this work are likely still a small subset of what one can learn from multi-messenger observations of wet EMRIs.  We believe there is significant scientific potential in studying joint science cases for wet EMRIs with GW and/or X-ray, optical and radio observations, as well as multiband events coming from stellar-mass binary mergers within AGN disks.


{\it\bf Acknowledgements---} 
We thank Jian-Ming Wang and Lianggui Zhu for interesting discussions. Zhenwei Lyu is supported by “the Fundamental Research Funds for the Central Universities” and Leicester International Institute, Dalian University of Technology.
This research has made use of the NASA/IPAC Extragalactic Database, which is funded by the National Aeronautics and Space Administration and operated by the California Institute of Technology.\\

{\it\bf Data Availability---}
The data that support the findings of this article are openly available at \url{https://github.com/gwlyu/wet_EMRIs.git}.\\

\appendix 
\section{Derivation of Sky Localization Volume}\label{sec:uncert}
\subsection{Fisher-matrix (quadratic) approximation}
In the main text, the detector response is modeled using two orthogonal TDI channels
$\alpha\in\{I,II\}$. For stationary Gaussian noise, the log-likelihood for parameters
$\boldsymbol{\lambda}=\{\lambda_i\}$ is
\begin{equation}
\ln \mathcal{L}(\boldsymbol{\lambda})
= -\frac{1}{2}\sum_{\alpha\in\{I,II\}}
\left\langle s_\alpha-h_\alpha(\boldsymbol{\lambda}),\, s_\alpha-h_\alpha(\boldsymbol{\lambda})\right\rangle\ ,
\label{eq:A1_loglike}
\end{equation}
where $s_\alpha$ is the (mock) data, $h_\alpha(\boldsymbol{\lambda})$ is the waveform template, and
$\langle\cdot,\cdot\rangle$ is the noise-weighted inner product defined in Eq.~(\ref{eq:inner}).

Let $\hat{\boldsymbol{\lambda}}$ denote the maximum-likelihood (ML) point and define offsets
$\Delta\lambda_i\equiv \lambda_i-\hat{\lambda}_i$. Expanding each channel waveform to first order about
$\hat{\boldsymbol{\lambda}}$ gives
\begin{equation}
h_\alpha(\boldsymbol{\lambda})
\simeq h_\alpha(\hat{\boldsymbol{\lambda}})
+ \sum_i \left.\frac{\partial h_\alpha}{\partial \lambda_i}\right|_{\hat{\boldsymbol{\lambda}}}\Delta\lambda_i\ ,
\label{eq:A2_taylor}
\end{equation}
where the sum runs over model parameters.

Defining the residual at the ML point in each channel,
\begin{equation}
r_\alpha \equiv s_\alpha - h_\alpha(\hat{\boldsymbol{\lambda}})\ .
\end{equation}
Using Eq.~(\ref{eq:A2_taylor}), the residual away from the ML point is
\begin{equation}
s_\alpha-h_\alpha(\boldsymbol{\lambda})
\simeq r_\alpha -
\sum_i \left.\frac{\partial h_\alpha}{\partial \lambda_i}\right|_{\hat{\boldsymbol{\lambda}}}\Delta\lambda_i\ .
\end{equation}
Substituting into Eq.~(\ref{eq:A1_loglike}) and expanding yields
\begin{widetext}
\begin{align}
\ln \mathcal{L}(\boldsymbol{\lambda})
&\simeq -\frac{1}{2}\sum_\alpha
\left[
\langle r_\alpha,r_\alpha\rangle
-2\sum_i \left\langle r_\alpha,\left.\frac{\partial h_\alpha}{\partial \lambda_i}\right|_{\hat{\boldsymbol{\lambda}}}\right\rangle \Delta\lambda_i + \sum_{i,j}
\left\langle
\left.\frac{\partial h_\alpha}{\partial \lambda_i}\right|_{\hat{\boldsymbol{\lambda}}},
\left.\frac{\partial h_\alpha}{\partial \lambda_j}\right|_{\hat{\boldsymbol{\lambda}}}
\right\rangle
\Delta\lambda_i\Delta\lambda_j
\right]\ ,
\label{eq:A2b_expand}
\end{align}
\end{widetext}

The linear term vanishes because the score is zero at the ML point, as follow:
\begin{equation}
0=\left.\frac{\partial \ln\mathcal{L}}{\partial \lambda_i}\right|_{\hat{\boldsymbol{\lambda}}}
=\sum_{\alpha}
\left\langle r_\alpha,\left.\frac{\partial h_\alpha}{\partial \lambda_i}\right|_{\hat{\boldsymbol{\lambda}}}\right\rangle\ .
\end{equation}
Subtracting the ML value
$\ln\mathcal{L}(\hat{\boldsymbol{\lambda}})=-(1/2)\sum_\alpha\langle r_\alpha,r_\alpha\rangle$ then gives the
standard quadratic (Fisher) approximation,
\begin{equation}
-2\ln\frac{\mathcal{L}(\boldsymbol{\lambda})}{\mathcal{L}(\hat{\boldsymbol{\lambda}})}
\simeq
\sum_{i,j}\Delta\lambda_i\,\Gamma_{ij}\,\Delta\lambda_j\ ,
\end{equation}
where the Fisher information matrix is
\begin{equation}
\Gamma_{ij}\equiv
\sum_{\alpha\in\{I,II\}}
\left\langle
\left.\frac{\partial h_\alpha}{\partial \lambda_i}\right|_{\hat{\boldsymbol{\lambda}}},
\left.\frac{\partial h_\alpha}{\partial \lambda_j}\right|_{\hat{\boldsymbol{\lambda}}}
\right\rangle \ .
\end{equation}
If the prior is approximately flat across the region of significant likelihood, the posterior is locally
Gaussian with covariance
\begin{equation}
\Sigma_{ij} \simeq (\Gamma^{-1})_{ij} \ .
\end{equation}
The off-diagonal elements $\Sigma_{ij}$ encode covariances (and hence correlations) between parameters $\lambda_i$ and $\lambda_j$, $\mathrm{Cov}(\lambda_i,\lambda_j)=\Sigma_{ij}$, while the diagonal elements give the marginalized variances, $\mathrm{Var}(\lambda_i)=\Sigma_{ii}\equiv \sigma_{\lambda_i}^2$.

\subsection{Joint $p$-probability ellipsoid in $N$ dimensions}\label{app:ellipsoid}
Consider an $N$-dimensional subset of parameters $\boldsymbol{\vartheta}$ (e.g.\ $\boldsymbol{\vartheta}=(\ln r,\theta,\phi)$)
with covariance sub-matrix $\Sigma_{\boldsymbol{\vartheta}}$ (the corresponding sub-block of $\Sigma_{ij}$). Define the Mahalanobis quadratic form
\begin{equation}
Q \equiv \Delta\boldsymbol{\vartheta}^{\mathsf T}\,\Sigma_{\boldsymbol{\vartheta}}^{-1}\,\Delta\boldsymbol{\vartheta}\ .
\end{equation}
If $\Delta\boldsymbol{\vartheta}$ is multivariate Gaussian with covariance $\Sigma_{\boldsymbol{\vartheta}}$, then
\begin{equation}
Q \sim \chi^2_{N}\ ,
\end{equation}
The \emph{joint} $p$-probability region is the ellipsoid
\begin{equation}
\mathcal{E}_p \equiv \left\{\Delta\boldsymbol{\vartheta}:\; Q \le A_{p,N}^2\right\}\ ,
\label{eq:A_ellipsoid}
\end{equation}
where $A_{p,N}^2$ is fixed by the $\chi^2_N$ quantile, as follow:
\begin{equation}
p = P(\chi^2_N \le A_{p,N}^2) =
\frac{\gamma\!\left(\frac{N}{2},\,\frac{A_{p,N}^2}{2}\right)}{\Gamma\!\left(\frac{N}{2}\right)}\ .
\end{equation}
Here $\gamma(\cdot,\cdot)$ and $\Gamma(\cdot)$ are the lower incomplete and complete $\Gamma$ functions (not to be confused with the Fisher matrix $\Gamma_{ij}$). 

Equivalently, one may write
\begin{equation}
A_{p,N}^2 = \chi^2_{N,p}\ ,
\end{equation}
where $\chi^2_{N,p}$ denotes the $p$-quantile of the $\chi^2$ distribution with $N$ degrees of freedom.
For the commonly used $p=0.6827$ joint credible regions,
\begin{align}
N=2:\quad & A_{0.68,\,2}^2=\chi^2_{2,\,0.68}\simeq 2.296\ ,
\; A_{0.68,\,2}\simeq 1.515\ ,\\
N=3:\quad & A_{0.68,\,3}^2=\chi^2_{3,\,0.68}\simeq 3.527\ ,
\; A_{0.68,\,3}\simeq 1.878\ .
\end{align}

\subsection{Hypervolume of the ellipsoid and numerical prefactors}\label{app:volume}
To compute the hypervolume enclosed by Eq.~(\ref{eq:A_ellipsoid}), we apply a whitening transformation.
Let $\Sigma_{\boldsymbol{\vartheta}} = LL^{\mathsf T}$ be a Cholesky factorization and define
\begin{equation}
\mathbf{y} \equiv L^{-1}\Delta\boldsymbol{\vartheta}\ .
\end{equation}
Then,
\begin{equation}
Q=\Delta\boldsymbol{\vartheta}^{\mathsf T}\Sigma_{\boldsymbol{\vartheta}}^{-1}\Delta\boldsymbol{\vartheta}
= \mathbf{y}^{\mathsf T}\mathbf{y}=\|\mathbf{y}\|^2\ ,
\end{equation}
so $\mathcal{E}_p$ maps to the $N$-ball $\|\mathbf{y}\|\le A_{p,N}$. The Jacobian determinant is
\begin{equation}
\left|\frac{\partial\Delta\boldsymbol{\vartheta}}{\partial\mathbf{y}}\right|
= |\det L| = \sqrt{\det\Sigma_{\boldsymbol{\vartheta}}}\ .
\end{equation}
Hence the $p$-probability ellipsoid hypervolume is
\begin{equation}
V_p^{(N)} =
V_{\mathrm{ball},N}(A_{p,N})\;\sqrt{\det\Sigma_{\boldsymbol{\vartheta}}}\ ,
\end{equation}
where the $N$-ball volume of radius $A$ is
\begin{equation}
V_{\mathrm{ball},N}(A)=\frac{\pi^{N/2}}{\Gamma\!\left(\frac{N}{2}+1\right)}\,A^N\ .
\end{equation}

In particular, the $N$-ball volumes are
\begin{equation}
V_{\mathrm{ball},2}(A)=\pi A^2 \quad\text{(area)}\ , 
\qquad
V_{\mathrm{ball},3}(A)=\frac{4\pi}{3}A^3 \quad\text{(volume)}\ .
\end{equation}
For the joint $p=0.6827$ region, this yields
\begin{align}
N=2:\;
& V_{0.68}^{(2)} = \pi A_{0.68,2}^2\,\sqrt{\det\Sigma_{\boldsymbol{\vartheta}}}
\;\simeq\; 7.21\,\sqrt{\det\Sigma_{\boldsymbol{\vartheta}}}\ ,\\[6pt]
N=3:\; 
& V_{0.68}^{(3)} = \frac{4\pi}{3}A_{0.68,3}^3\,\sqrt{\det\Sigma_{\boldsymbol{\vartheta}}}
\;\simeq\; 27.74\,\sqrt{\det\Sigma_{\boldsymbol{\vartheta}}}\ .
\end{align}

\subsection{Sky solid angle and 3D localization volume}\label{app:sky_and_volume}
We specialize the joint $p$-ellipsoid construction to (i) sky angles and (ii) the 3D localization volume.

{\it Solid-angle (sky) uncertainty.}
Let $\boldsymbol{\vartheta}=(\theta,\phi)$ with covariance sub-matrix
$\Sigma_{\theta,\phi}$ (the $2\times2$ sub-block of $\Sigma_{ij}$ evaluated at
$\hat{\boldsymbol{\lambda}}$). The area of the joint $p$ ellipse in the
\emph{coordinate} plane $(\theta,\phi)$ is
\begin{equation}
V_p^{(2)} \;=\; \pi A_{p,2}^2\,\sqrt{\det\Sigma_{\theta,\phi}}\ ,
\qquad
A_{p,2}^2=\chi^2_{2,p}\ .
\end{equation}
The physical solid-angle element on the unit sphere is
\begin{equation}
d\Omega=\sin\theta\,d\theta\,d\phi\ .
\end{equation}
If the uncertainty region is sufficiently small that $\sin\theta$ may be treated as constant across it, evaluated at the best-fit sky location $\hat{\theta}$, then the joint $p$ solid-angle uncertainty is
\begin{equation}
\Delta\Omega_p \;\simeq\; \sin\hat{\theta}\;V_p^{(2)}
\;=\; \pi A_{p,2}^2\,\sin\hat{\theta}\,\sqrt{\det\Sigma_{\theta,\phi}}\ .
\end{equation}
For $p=0.6827$, $A_{0.68,\,2}\simeq 1.515$
\begin{equation}
\boxed{\Delta\Omega_{0.68}
\simeq 7.21\,\sin\hat{\theta}\,\sqrt{\det\Sigma_{\theta,\phi}}}\ .
\end{equation}

{\it Physical sky-localization volume.}
For 3D localization take $\boldsymbol{\vartheta}=(\ln r,\theta,\phi)$ with
covariance sub-matrix $\Sigma_{\ln r,\theta,\phi}$. The physical volume element is
\begin{equation}
dV = r^2\sin\theta\,dr\,d\theta\,d\phi
= r^3\sin\theta\,d(\ln r)\,d\theta\,d\phi\ .
\label{eq:A_dV}
\end{equation}
Over a sufficiently small region around the best-fit parameters, we approximate the Jacobian factor $r^3\sin\theta$ as constant and evaluate it at $(\hat{r},\hat{\theta})$. The joint $p$ ellipsoid in $(\ln r,\theta,\phi)$ has parameter-space hypervolume
\begin{equation}
V_p^{(3)} \;=\; \frac{4\pi}{3}A_{p,3}^3\,
\sqrt{\det\!\left(\Sigma_{\ln r,\theta,\phi}\right)}\ .
\qquad
A_{p,3}^2=\chi^2_{3,p}\ .
\end{equation}
Multiplying by the physical Jacobian factor from Eq.~(\ref{eq:A_dV}) gives the physical localization volume
\begin{equation}
\Delta V_p \;\simeq\;
\hat{r}^{\,3}\sin\hat{\theta}\;
\frac{4\pi}{3}A_{p,3}^3\,
\sqrt{\det\!\left(\Sigma_{\ln r,\theta,\phi}\right)}\ .
\label{eq:A_deltaV_p}
\end{equation}
For $p=0.6827$, $A_{0.68,\,3}\simeq 1.878$, hence
\begin{equation}
\boxed{\Delta V_{0.68}
\simeq27.74\,\hat{r}^{\,3}\sin\hat{\theta}\,
\sqrt{\det\!\left(\Sigma_{\ln r,\theta,\phi}\right)}}\ .
\end{equation}
This expression accounts for correlations among $(\ln r,\theta,\phi)$ through the determinant of the full $3\times3$ covariance sub-matrix.

\subsection{Remark on factorized approximations}\label{app:factorized}
A common shortcut is to ``factorize'' the 3D localization volume into a radial part and a sky part, as follows:
\begin{equation}
\Delta V \approx \hat{r}^3\,\sigma_{\ln r}\,\Delta\Omega\ .
\end{equation}
This approximation is valid only if: (i) $\ln r$ is (nearly) uncorrelated with the sky angles $(\theta,\phi)$, so the 3D covariance is approximately block-diagonal; and (ii) $\sigma_{\ln r}$ and $\Delta\Omega$ are defined so that they correspond to the same \emph{joint}
probability content $p$.

In general the posterior in $(\ln r,\theta,\phi)$ is a correlated 3D Gaussian (under the Fisher approximation), so the correct joint $p$ region is a 3D ellipsoid. Its physical volume for $p=0.6827$ is expressed as Eq.~(\ref{eq:A_deltaV_p}).
Therefore an expression such as
\begin{equation}
\Delta V_{0.68}= \hat{r}^3\,\sigma_{\ln r}\,\pi\sin\hat{\theta}\,\sqrt{\det(\Sigma_{\theta,\phi})}\ .
\end{equation}
is generally inconsistent: it (a) ignores correlations between distance and sky position, and (b) mixes a 1D and 2D uncertainty in a way that does not represent a \emph{joint} 68\% region; the correct joint-3D numerical factor is $\simeq 28$, not $\pi$.

\nocite{*}
\bibliography{ref}

\end{document}